\documentclass[10pt,twocolumn]{article}

\usepackage{nicematrix}
\usepackage{tabularray}

\usepackage{graphicx}
\usepackage{dcolumn}
\usepackage{amsmath}
\usepackage{authblk}
\usepackage{multirow}
\usepackage{array}
\usepackage{booktabs}
\usepackage{bm}
\usepackage{float}
\usepackage{braket}
\usepackage{xcolor}
\usepackage{wrapfig}
\usepackage{mathrsfs}
\usepackage[rightcaption]{sidecap}
\usepackage[super]{natbib}
\newtheorem{definition}{Definition}
\usepackage[hidelinks]{hyperref}

\usepackage[normalem]{ulem}
\newcommand{\stkout}[1]{\ifmmode\text{\sout{\ensuremath{#1}}}\else\sout{#1}\fi}

\newcommand{\elm}[3]{\left<#1\left|#2\right|#3\right>}
\begin{document}


\title{Quantum Information Driven Ansatz (QIDA): shallow-depth empirical quantum circuits from Quantum Chemistry}

\author[1,2]{Davide Materia}
\author[1,2] {Leonardo Ratini}
\author[3]{Celestino Angeli}
\author[1]{Leonardo Guidoni
\thanks{leonardo.guidoni@univaq.it}}

\affil[1]{%
Dipartimento di Scienze Fisiche e Chimiche, Universit\`a degli Studi dell’Aquila, Coppito, L’Aquila, Italy }%

\affil[2]{Dipartimento di Ingegneria e Scienze dell'Informazione e Matematica\\ Universit\`a degli Studi dell'Aquila, Coppito, L'Aquila, Italy}

\affil[3]{Dipartimento di Scienze Chimiche, Farmaceutiche ed Agrarie, Università degli Studi di Ferrara, Italy}

\date{\today}

\twocolumn[
\begin{@twocolumnfalse}
\maketitle
\begin{abstract}
Hardware-efficient empirical variational ans\"atze for Variational Quantum Eigensolver simulations of Quantum Chemistry suffer from the lack of a direct connection to classical Quantum Chemistry methods. In the present work, we propose a method to fill this gap by introducing a new approach for constructing variational quantum circuits, leveraging quantum mutual information associated with classical Quantum Chemistry states to design simple yet effective heuristic ans\"atze with a topology that reflects the correlations of the molecular system.
As first step, Quantum Chemistry calculations, such as Møller-Plesset (MP2) perturbation theory, firstly provide an approximate Natural Orbitals basis, which has been recently shown to be the best candidate one-electron basis for developing compact empirical wavefunctions (Ratini, et al 2023). Secondly, throughout the evaluation of quantum mutual information matrices, they provide information about the main correlations between qubits of the quantum circuit, enabling the development of a direct design of entangling blocks for the circuit. The resulting ansatz is then utilized with a Variational Quantum Eigensolver (VQE) to obtain a short depth variational groundstate of the electronic Hamiltonian. To validate our approach, we perform a comprehensive statistical analysis by simulations over various molecular systems ($H_2, LiH, H_2O$) and apply it to the more complex $NH_3$ molecule.
The reported results demonstrate that the proposed methodology gives rise to highly effective ansätze, surpassing the standard empirical ladder-entangler ansatz in performance. Overall, our approach can be used as effective state preparation providing a promising route for designing efficient variational quantum circuits for large molecular systems.

\end{abstract}
\vspace{1cm}
\end{@twocolumnfalse}

]

\section{\label{sec:Intro}Introduction}

The advent of quantum computers promises to drastically reduce the computational cost of specific tasks with respect to their classical counterparts. With applications ranging from cryptography\cite{Shor1997} to finance\cite{Egger2021}, the public interest in the field has in recent years grown immensely\cite{Seskir2022}, particularly for its academic applications in the field of natural sciences.\cite{Troyer}\cite{Robert2021} In line with one of the precursors proposals of quantum computing \cite{Feynman2018}, one of the most promising fields is thought to be Quantum Chemistry, which aims to tackle the electronic structure of large molecules where electron correlation plays a crucial role.

A widely known method for near-term applications in quantum chemistry is the Variational Quantum Eigensolver (VQE)\cite{Peruzzo2014, Bharti2022, Cerezo2021}, used to variationally approximate the groundstate of a molecular Hamiltonian by routinely measuring on the Quantum Processing Unit (QPU) the energy of a parameterized quantum circuit, i.e. the wavefunction ansatz.

Different approaches to building an ansatz have been developed. Among all of them, two major classes can be found. The first consists in translating classical Quantum Chemistry approaches \footnotetext{in the sense of not arising from quantum computing} into the language of quantum computation. One of the most representative  examples is the
Unitary Coupled Cluster (UCC) method\cite{Hoffmann1998, Kutzelnigg1991, Cooper2010, Evangelista2011, Whitfield2011, Barkoutsos2018, Romero2018} which is a unitary implementation of the widely used Coupled Cluster classical method. The second approach, on the contrary, starts from wavefunctions that are directly built exploiting the characteristic of the quantum hardware (for instance, by reducing as much as possible the depth of the quantum circuit), regardless of connections with the Chemistry of the problem. This empirical approach defines the so-called Heuristic Ansatz\cite{Kandala2017, rattew2020, Tang2021, Ratini2022}, which is composed of repetitions of blocks of rotations and entanglements, set up without any information of the chemical system.
Compared to the former approach, this one better exploits the capabilities of the quantum hardware, at the price of losing the chemical meaning of the variational ansatz.

The aim of our work is to join these two approaches by building, within the realm of the latter method, short-depth circuits integrating knowledge derived by Classical albeit affordable Quantum Chemistry calculations such as second order Møller-Plesset perturbation theory (MP2). \cite{mp2}
The key physical quantity for this bridge is the Quantum Mutual Information (QMI). In general Quantum Chemistry, the Quantum Mutual Information has already appeared under different aspects, its main objective however has always been to give a measure of correlation between different orbitals.~\cite{Ding2021} During the years, this tool has been used to help converge DMRG calculations~\cite{Legeza2003, Rissler2006} or to establish a black-box method of selecting active orbitals within the CASSCF framework.~\cite{Stein2016} In quantum computing, QMI
has already been used for suggesting the best logical-to-physical assignment for the qubits~\cite{Tkachenko2021} or improving existing algorithms.~\cite{Zhang2020}
In our approach, the Quantum Information Driven Ansatz (QIDA) is built with a circuit topology that reflects the QMI derived by preliminary quantum chemistry calculations as will be detailed in the next sections. This ansatz represents a compact and affordable starting point on which to construct more complex wavefunctions.
Together with such analysis, a further enhancement of the results offered by our scheme will be also provided by the use of Natural Orbitals (NO) as the preferred basis where developing correlated compact wavefunctions, as recently shown by us in the context of WAHTOR algorithm. \cite{Ratini2022}
The background of quantum mutual information-aided VQE makes this work complementary to the approach reported in ref. \cite{Zhang2020}, which treats a similar aim in the context of a UCC ansatz within an adaptive framework.\cite{Grimsley2019} \\
In the following sections, inspired by the work of Tkanchenko~et~$al$~\cite{Tkachenko2021}, we introduce the Quantum Information Driven Ansatz (QIDA) method, applying it to the evaluation of ground state energies of representative benchmark molecules such as $H_2$, $LiH$, $H_2O$, and $NH_3$.

\section{\label{sec:QIDA}Quantum Information Driven Ansatz (QIDA)}
\subsection{Mutual-Information Metrics}
The measure of correlations in quantum systems can be achieved in many ways, and nowadays it constitutes a field of research by itself. \cite{Amico2008} One of the most prominent and oldest measures of correlation is the Von-Neumann quantum mutual information.\cite{Neumann}

\begin{definition}
Let $\mathscr{H}_A$ and $\mathscr{H}_B$ be two Hilbert spaces and $\rho_{AB}$ be a density matrix, 
the quantum mutual information $I$ of $\rho_{AB}$ is defined as: 
\begin{equation}
    I(A,B)=S(A)+S(B)-S(A,B)
    \label{eq:qmi}
\end{equation}
    where $S$ are the Von Neumann entropies defined as
\begin{equation}
\begin{split}    S(A,B)=-tr(\rho_{AB}log(\rho_{AB}))\qquad\\
    S(A)=-tr(\rho_{A}log(\rho_{A}))\qquad\quad\\
    \rho_{A}=tr_{B}(\rho_{AB})\qquad\qquad
\end{split}
\label{eq:entropy}
\end{equation}
\end{definition} 

As shown by equation~\eqref{eq:entropy}, given an N-partite Hilbert space $\mathscr{H}_N=  \bigotimes_i^N\mathscr{H}_i$ and a quantum state $\rho \in \mathscr{H}_N$, we can find the matrix $I(i,j)$ by tracing out all but the subsystems $i,j$. In our case, subsystems are $N$ qubits forming the quantum register. By construction, the matrix is symmetric and the elements along the diagonal must be zero.

The use of the QMI for our purpose was motivated by its simplicity and widespread knowledge in both the fields of quantum and classical information theory. Furthermore, the QMI has already been exploited to measure the correlation between orbitals in the past.\cite{Legeza2003, Rissler2006}
 
We will later show the properties of $I$ matrix through representation such as figure \ref{fig:maps}.

\subsection{Natural orbitals}
\label{subsec:NO}
In the following sections, we make use of NO as a tool to decrease the complexity of the circuit, using them as one-electron basis sets for VQE calculations as well as for QMI calculations.

The NO of a given wave function $\Psi$ are defined as the molecular orbitals for which the one-body Reduced Density Matrix (RDM) 
\begin{equation}
     \rho_{ij}=\elm{\Psi}{a^{\dag}_{i}a_{j}}{\Psi}
\end{equation}
 is diagonal. Due to this diagonal condition, the NO allow a more classical interpretation of the electrons' placement over the orbitals, as the element $\rho_{ii}$ gives an
indication of the number of electrons in the $i$-th orbital and is called the occupation number of orbital $i$.

\begin{figure*}[]
    a)
    \includegraphics[width=.270\textwidth]{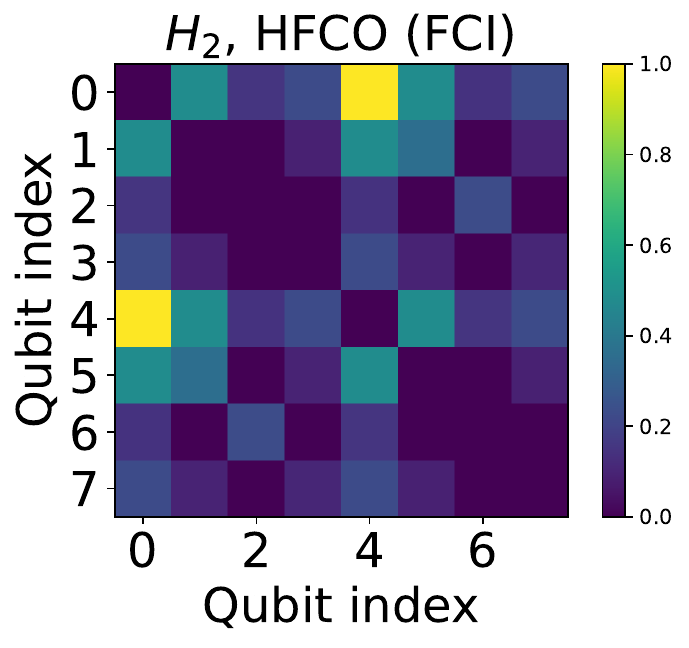}\hfill
    b)
    \includegraphics[width=.270\textwidth]{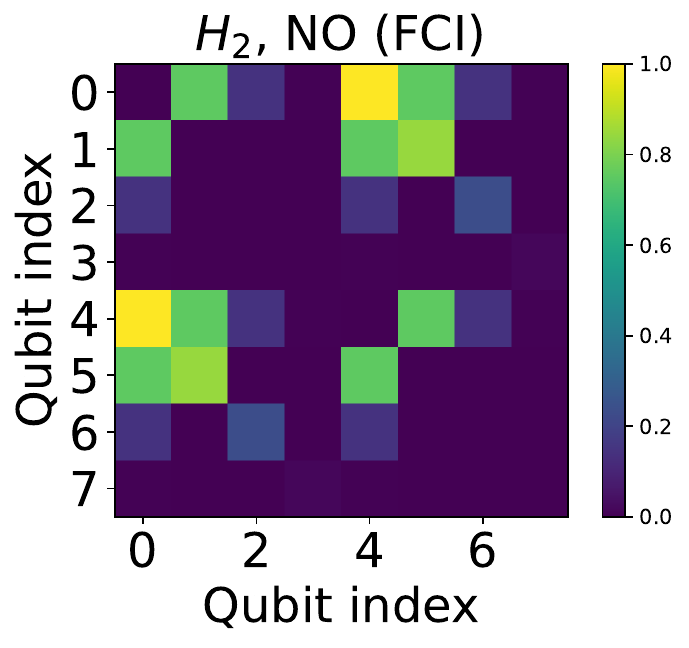}\hfill
    c)
    \includegraphics[width=.270\textwidth]{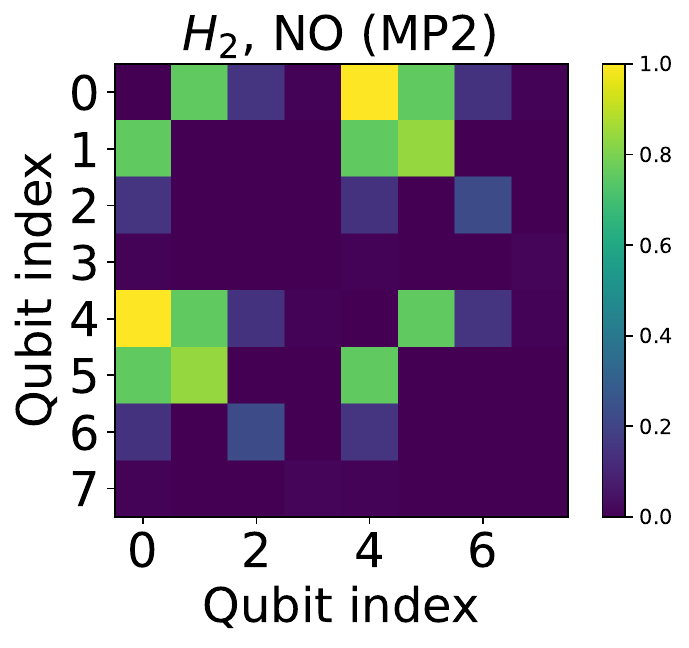}
    \\[\smallskipamount]
    d)
    \includegraphics[width=.270\textwidth]{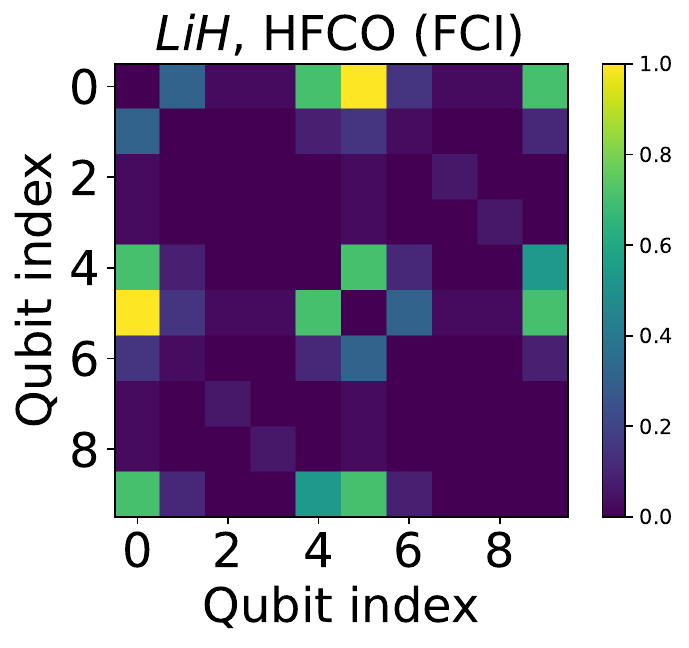}\hfill
    e)
    \includegraphics[width=.270\textwidth]{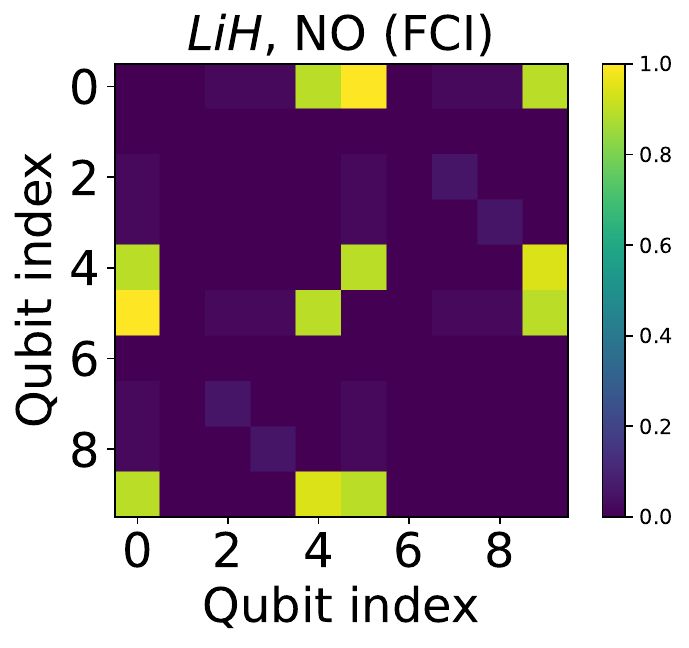}\hfill
    f)
    \includegraphics[width=.270\textwidth]{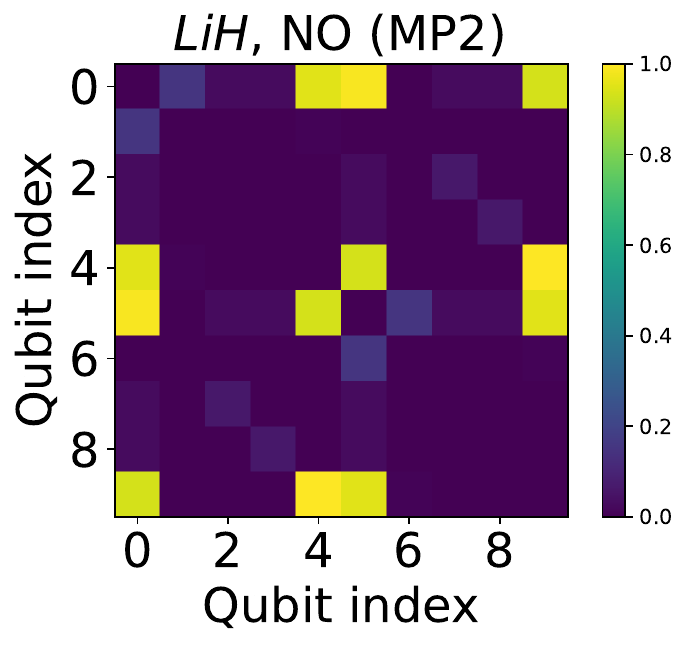}
    \\[\smallskipamount]
    h)
    \includegraphics[width=.270\textwidth]{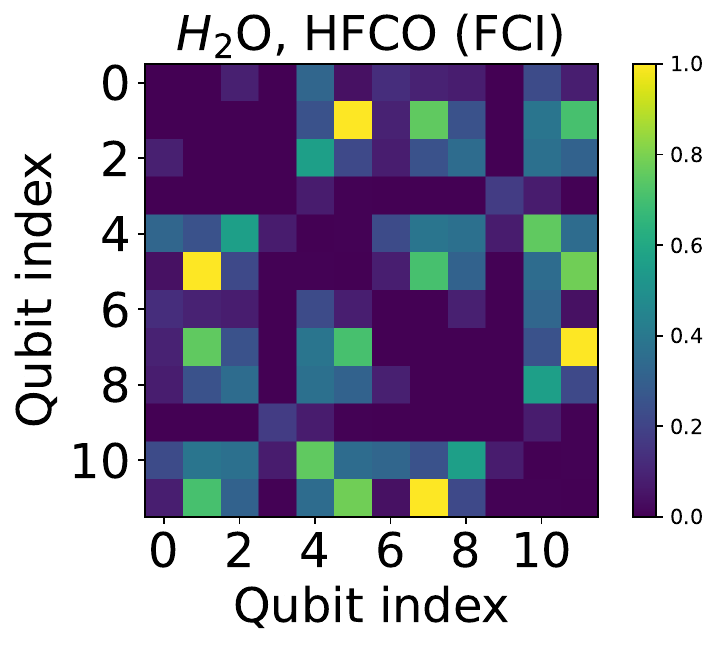}\hfill
    i)
    \includegraphics[width=.270\textwidth]{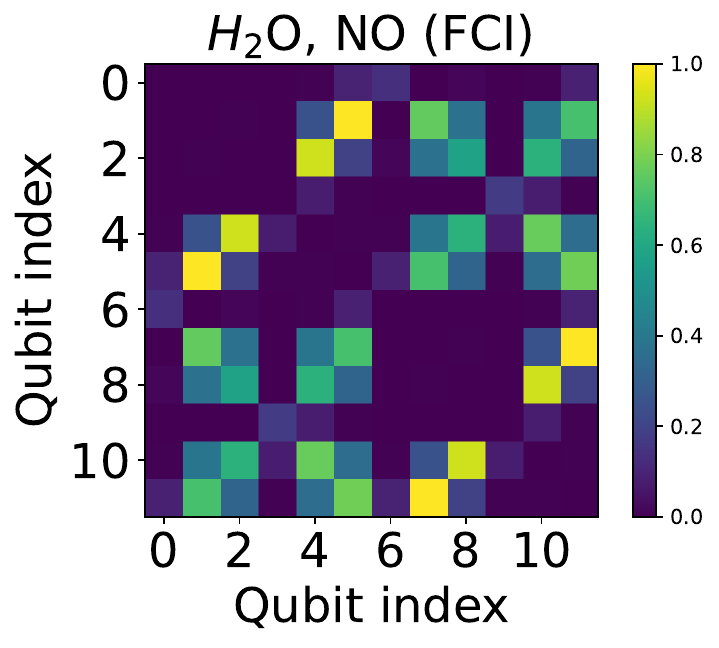}\hfill
    j)
    \includegraphics[width=.270\textwidth]{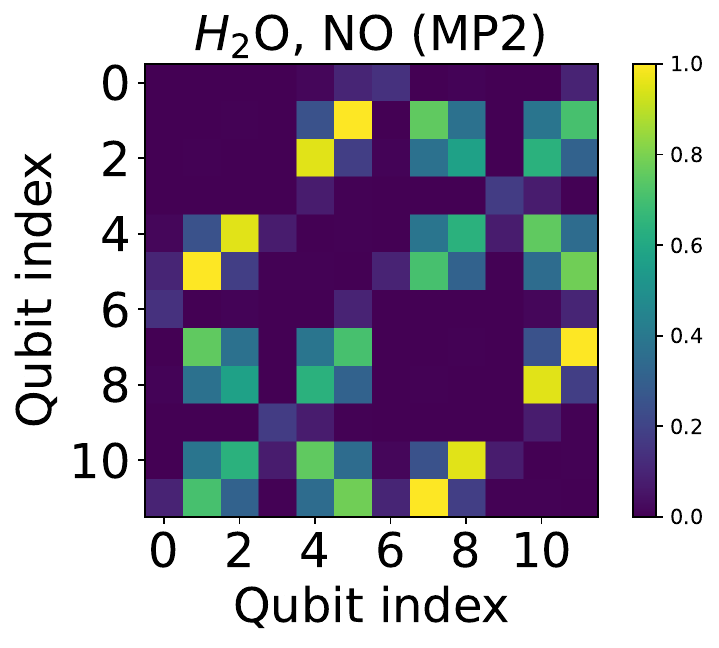}
    \\[\smallskipamount]
    k)
    \includegraphics[width=.270\textwidth]{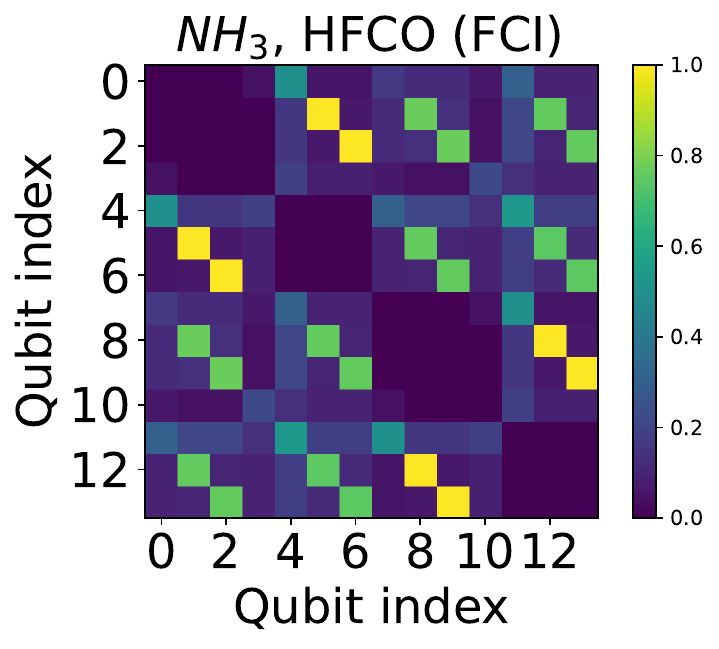}\hfill
    l)
    \includegraphics[width=.270\textwidth]{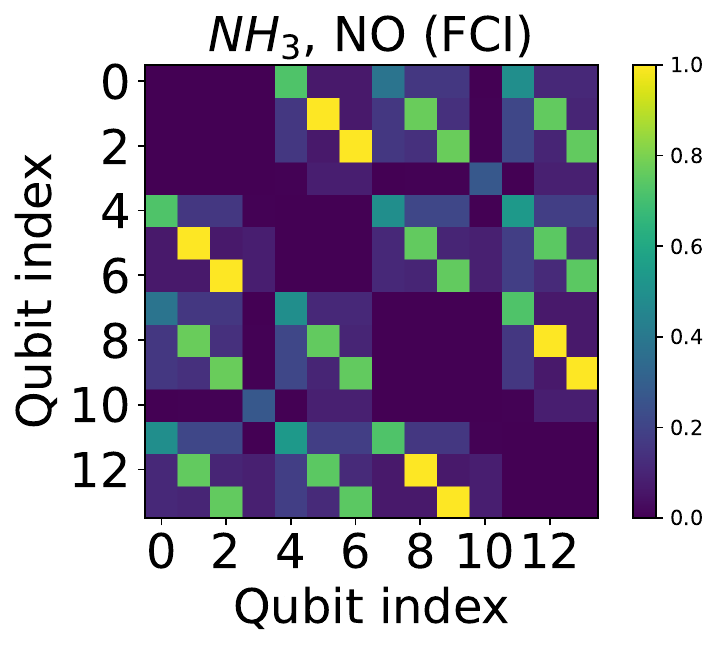}\hfill
    m)
    \includegraphics[width=.270\textwidth]{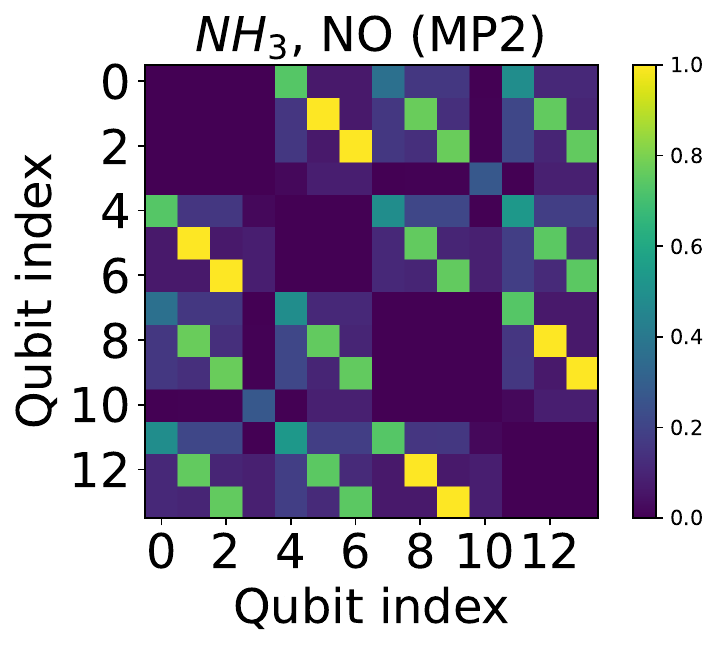}
    \caption{Representation of the quantum mutual information matrices $I(i,j)$ for qubits mapping the spin-orbitals of the molecular systems considered. The rows span over the molecules, while the columns span over the orbitals and the states: starting from the left we meet FCI state in HFCO for the first column, and then FCI and MP2 states each in its respective NO for the second and third column.}
    \label{fig:maps}
\end{figure*}

Furthermore, it has been claimed by Löwdin~\cite{lowdin} that, in the NO basis, the configuration interaction expansion of the state under study is expressed with the minimal number of Slater determinants. 
Consequently, we can hope to see a sparser $I$ matrix by replacing the Hartree-Fock~\cite{Hartree1928} (HF) canonical Orbitals (HFCO) with the NO of a correlated wave function. It is worth recalling here that the HF canonical orbitals satisfy the HF energy minimization condition and diagonalize the Fock matrix. 
This property has been empirically verified by us for prototypical molecules in a previous work by comparing the $I$ matrix for an heuristic wave function based on HF canonical orbitals with that based on optimized orbitals, which turned out to be always very close to the NO for all cases of study. \cite{Ratini2023}
This behavior can be also seen in image \ref{fig:maps}, representing the $I$ matrix for the set of benchmarking molecules taken into account in this work. For each row, by moving from columns 1(a,d,g, HFCO) to columns 2(b,e,h, NO), the number of colored spots decreases in favor of some dominating ones, showing that the correlation concentrates more on fewer pairs of qubits. The usefulness of this feature will become clear in the following.
As previously reported, a set of NO is defined by diagonalization of the RDM of a state. In the case of proper NO the state would be the ground state, thus in principle implying prior knowledge of the exact Full Configuration Interaction~\cite{Knowles1984, Szabo1989} (FCI) solution. To overcome this problem, we consider an approximated set of NO obtained by the diagonalization of the RDM obtained by MP2 calculation. \cite{mp2} 

\subsection{The scheme}
\label{subsec:Scheme}
Resource-efficient classical quantum chemistry methods can already contain useful information for a quantum computing algorithm to solve electronic problems. This is not a new concept as indeed the starting point for VQE calculations is often the HFCO basis, even if starting from some other basis is possible.

In this work we developed a scheme that makes use of the QMI coming from a classical calculation to infer an entangling block for a hardware-efficient ansatz.

 \begin{figure}[h]
    \hspace{0.03\textwidth}
    \includegraphics[width=0.4\textwidth]{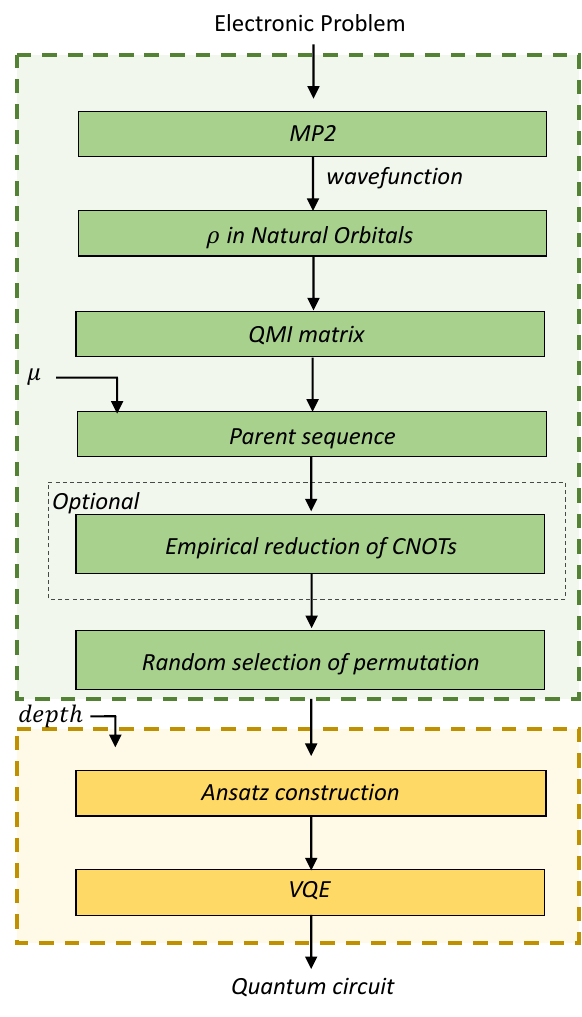}
    \caption{Description of the proposed scheme to obtain an ansatz from information inferred by a post-Hartree-Fock method, MP2 in this work.}
    \label{fig:scheme}
\end{figure}

As a first step, we perform a Post-HF (PHF) calculation to obtain an approximation of the GS $\ket{\psi_{PHF,GS}}$. The choice of the method is usually dictated by the size of the problem. We perform a second-order Møller-Plesset (MP2)\cite{mp2} calculation which scales with $O(n^5)$\cite{Szabo1989} in the number of orbitals.
As a following step, we map \cite{Jordan1928, Bravyi2002, Parity} the state $\ket{\psi_{PHF,GS}}$ on a qubit register to obtain $\ket{\psi_{q,PHF,GS}}$. 

At this point, we find the corresponding $I$ matrix by using equation \ref{eq:qmi}. Some explicit examples of these matrices are given in fig~\ref{fig:maps}, with elements ranging from 0 to 1 because of enforced normalization. \\
Overall, we can consider these values of I as a measure of correlation~\cite{Zeng2019}, without discerning between the classical correlation and the quantum correlation.\\
We can therefore identify which pairs have a larger contribution to this correlation by fixing a threshold $\mu \in [0,1)$, dividing our matrix into two separate sets of the pairs falling either above or below $\mu$. These data are used to design our empirical ansatz, by
considering the entangling block of the VQE to be composed of CNOTs only between qubits $i,j$ for which $I(i,j)>\mu$, leaving aside the direct imposition of the correlation of the remaining pairs or assigning it to further development of the ans\"atz.
In this framework, coherently to usual heuristic approaches \cite{Ratini2022}, the CNOT is chosen as an entangler (i.e. correlator) between qubits. Of course, other two-qubit quantum gates could also fulfill the same role.

The general idea is that a large value of $\mu$ selects the most relevant correlations between qubits. By decreasing the $\mu$ parameter, one can now choose in a consistent way how many CNOTs to include, to build more and more correlated ans\"atze, at the price of a longer circuit depth.
To further reduce the number of CNOTs of the ansatz we also introduce a special protocol that will be described in the next subsection.
Eventually, after having chosen a $\mu$ and having applied a CNOTs reduction method, one will have selected a list of CNOTs to be applied as an ans\"atz, named the \textit{parent} sequence. 

It is necessary to name such a sequence as the algorithm here proposed doesn't specify the order of the involved CNOTs, and since one has to entangle at least O(\textit{n}) qubits in a \textit{n} qubit register to achieve meaningful results, the number of possible orderings will grow as \textit{n}!. Thus in this context, the role of the just defined \textit{parent} sequence is to represent the set of all possible permutations generated from itself.
We show below that choosing a random permutation between the \textit{n}!, will still give better statistical results than the common ladder-entangler block with less or equal numbers of CNOTs. 

\subsubsection{Empirical reduction}
\label{subsubsec:reduction}

\begin{figure}[h]
    \centering     
    \includegraphics[width=.48\textwidth]{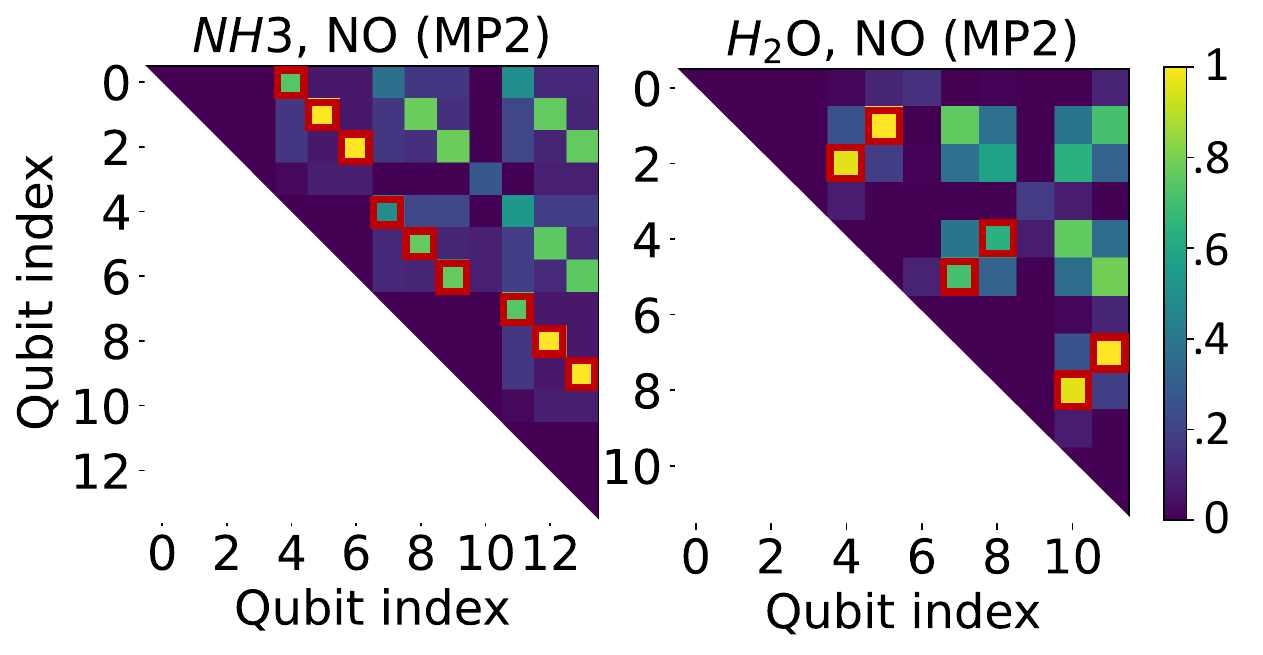}
    \caption{Empirical reduction criterion applied  to $H_2O$ and $NH_3$ molecules. As described in the main text, spots with red frames identify the qubits pairs that will be entangled with the two-qubit gate (CNOT in our case).}
    \label{fig:CNOTs_reduced}
\end{figure}

In order to retain the scaling of the two-qubit gate count of the ladder ansatz, bilinear in the number of qubits and depth, we found that it is convenient to remove some of the CNOTs belongign to groups which are cross-entangled, i.e. forming closed-loops of connections between different qubits.
Figure \ref{fig:maps} can help to clarify the concept of cross-entanglement. In this specific case all the pairs obtained from qubits 1,5,7 have an $I$ between each other greater than a certain $\mu$ threshold.
\begin{figure}[h]
    \centering
     a)
     \includegraphics[width=.225\textwidth]{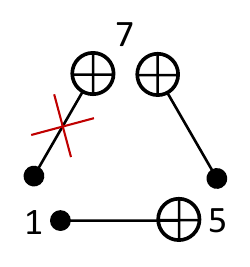}    
     b)
    \includegraphics[width=.2\textwidth]{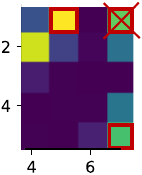}
    \caption{Cross-entanglement reduction. All numbers refer to the qubit indices. a) We have three CNOTs but we do not consider the one connecting the qubits 1 and 7. b) Representation on $I$ map of the reduction. The red frame identifies the pairs that we connect with a CNOT on the circuit whereas the crossed one is left out.}
    \label{fig:Triangolo}
\end{figure}
As described in the previous section, our procedure suggests to insert three CNOTs corresponding to all three pairs. Nevertheless, in a variational algorithm having multiple depths, it should be sufficient to connect these qubits only once. 
This means that it is possible to remove one of the three CNOTs, as represented in figure \ref{fig:Triangolo}[a]. We represent graphically this operation by putting a red frame on the considered spots and erasing the one that will not appear on the circuit, as done in figure \ref{fig:Triangolo}[b].
Starting from this consideration, we have implemented an empirical strategy in which only the first spot from each line of the upper triangular part of the $I$ matrix such that $I(i,j)>\mu$ is selected. As an example, figure \ref{fig:CNOTs_reduced} shows the pairs taken into account for the $H_2O$ and $NH_3$ molecules.

\section{\label{sec:details}Computational details}
\subsection{\label{sec:level2}Tested Systems}
The part of our algorithm undertaken by classical Quantum Chemistry approaches such as the HF and electronic integral calculations is performed with the PySCF python package\cite{pyscf}, whereas the MP2 wave functions are obtained by an adaptation of the MolCAs package.\cite{molcas}

The fraction of the correlation energy $\epsilon$ is defined by
\begin{equation}
\epsilon = \frac{E_{X}-E_{HF}}{E_{ex,c}}, \quad E_{ex,c} = E_{FCI}-E_{HF}
\end{equation}
where $E_{x}$ is the energy obtained from the variational calculation and $E_{HF}$ and $E_{FCI}$ are the Hartree-Fock and full configuration interaction energies.

The core software used for quantum computing simulations is the Qiskit Python library.\cite{Qiskit} The Jordan-Wigner \cite{Jordan1928} encoding method has been considered in each simulation to map the fermionic system on a qubit register. 

\begin{table*}[h]
\centering
\newcommand{\minitab}[2][l]{\begin{tabular}#1 #2\end{tabular}}
\begin{tabular}{|c | c | c | c| c| c|}
\hline
\hline 
Mol. &  Coordinates (\AA{}) &  Qubits & $E_{ex,c}${\tiny$(\frac{Kcal}{mol})$} & $\mu$ & \# elements in parent sequence \\ 
\hline
\hline
$H_2$   & \minitab[c]{H 0.0 0.0 0.0 \\ H 0.0 0.0 0.74}  & 8  & 15.636 & 0.5  & 6 \\
\hline
$LiH$   & \minitab[c]{Li 0.0 0.0 0.0 \\ H 0.0 0.0 1.595}  & 10  & 12.6446 & 0.5  & 6 \\
\hline 
$H_2O$  & \minitab[c]{O 0.0 0.0 0.0 \\ H 0.757 0.586 0.0 \\ H -0.757 0.586 0.0} 
& 12  & 31.006 & \minitab[c]{0.5  \\ 0.7 \\ 0.5 \textit{red.}} &  \minitab[c]{12 \\ 9 \\ 6}\\
\hline
$NH_3$  & \minitab[c]{N 0.0 0.0 0.1211 \\ H 0.0 0.9306 -0.2826 \\ H 0.8059 -0.4653 -0.2826	\\ H -0.8059 -0.4653 -0.2826}  & 14 & 41.287  & 0.5 \textit{red.} & 9 \\
\hline	
\end{tabular}
\caption{Summary of data of the systems under analysis and computational details. $H_2$ has been analyzed with 631-G basis while every other molecule used STO-3G basis set and frozen core approximation at the HF level}

\label{tab: systems}
\end{table*}

In table \ref{tab: systems} we list the data corresponding to the molecules under investigation.

\subsection{Construction of the Ansatz}
\label{subsec:Ansätze}
In section \ref{subsec:Scheme} we have illustrated how we can define parent sequences of CNOTs to be implemented as an ansatz on the quantum circuit. Anyway, given the arbitrary choice of CNOTs ordering in such sequences, we still have a factorial scaling if we would like to explore all possible orderings. To explore the effects of different choices, for short parent sequences, we have systematically explored all combinations; on the contrary, for parent sequences longer than 6 CNOTs the analysis has been taken on a restricted random selection of ordering. For the randomly chosen ans\"atze we preferred a uniform distribution for the qubits to be the target of a CNOT. Given the convention in the field of preferring top-down entanglement, the qubits we are referring range from the second to the last one. 
We proceed as following: we randomly extract $t$ from the set $\{1,\dots,n\}$ and $c$ from the set $\{0,\dots,p-1\}$, then we add the CNOT with control on qubit $c$ and target on qubit $t$ to the entangling block repeating the procedure until it contain $n-1$ CNOTs.
After every application of the chosen entangling block, we apply a $ R_{Y}(\theta)$ quantum gate on every qubit. These two sections of the circuit together are referred to as $block$ and with the term $depth$ we refer to the number of blocks that will be repeated to build the final ansatz. It is worth noticing that the variational parameters are independent for each block.
In the following we show the advantages of our QMI-aided ansatz over the ladder-entangler ansatz, by analyzing the results at different depths.
We also point out that the restricted number of entangling blocks suggested by our algorithm does not necessarily connect all the qubits, as also happens in the cases under analysis. 
With few exceptions, it is then reasonable to expect that a more expensive ladder-entangler block that links all the qubits will generally give the possibility to span the whole Hilbert space at the price of increasing the gate count. 

Table \ref{tab: systems} shows the values of the $\mu$ parameters investigated and the corresponding identified parent sequences. 
Concerning the choice of the $\mu$ parameter, since the systems under analysis are not yet big enough to show a clear lowering of the number of CNOTs as a function of $\mu$, we only consider one value for each system. An exception is considered for the $H_2O$ molecule, for which we used the two values $\mu = 0.5$ and $\mu = 0.7$. 
\begin{figure}[]
    \centering
    \includegraphics[width=0.35\textwidth]{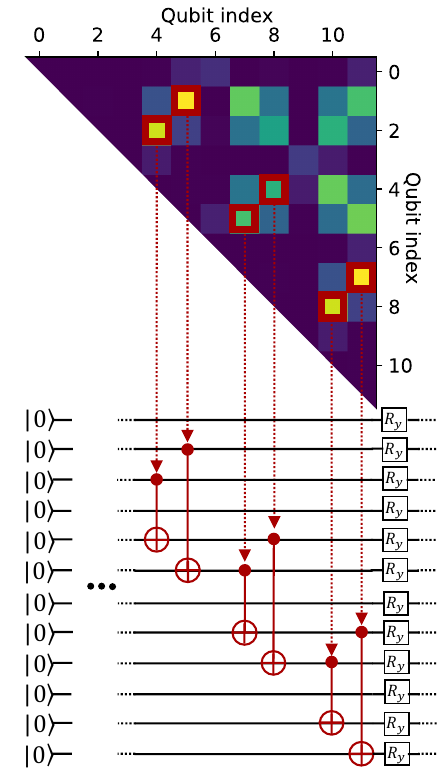}
    \caption{Pictorial representation of the method here considered for $H_2O$ molecule. The empirical reduction has been applied to the parent sequence obtained by $\mu=0.5$. The simple permutation represented in the picture is chosen purely for illustrative purposes.}
    \label{fig:arrows}
\end{figure}

An example of ansatz construction for the $H_2O$ molecule is given in figure \ref{fig:arrows}. It has been obtained by choosing one of the permutations of the parent sequence generated by setting $\mu=0.5$ and taking into account the empirical reduction.
For all the following plots we performed an analysis on 100 different possible permutations of the original parent sequence. Even for the random maps, we performed analysis on 100 of them. In all cases, we performed 50 VQE runs for each entangling block with random starting parameters to guarantee sufficient statistical value for each ans\"atze. The only exception is the ladder-entangler block, for which we performed 300 VQE since it represents the reference entangling block for comparison.
Given its relevance for the results discussed below, the depth 4 of the ladder-entangler ansatz of the water molecule has received 2000 repetitions.

\section{\label{sec:results}Results}
\begin{figure}[h!]
    \centering
    \includegraphics[width=0.47\textwidth]{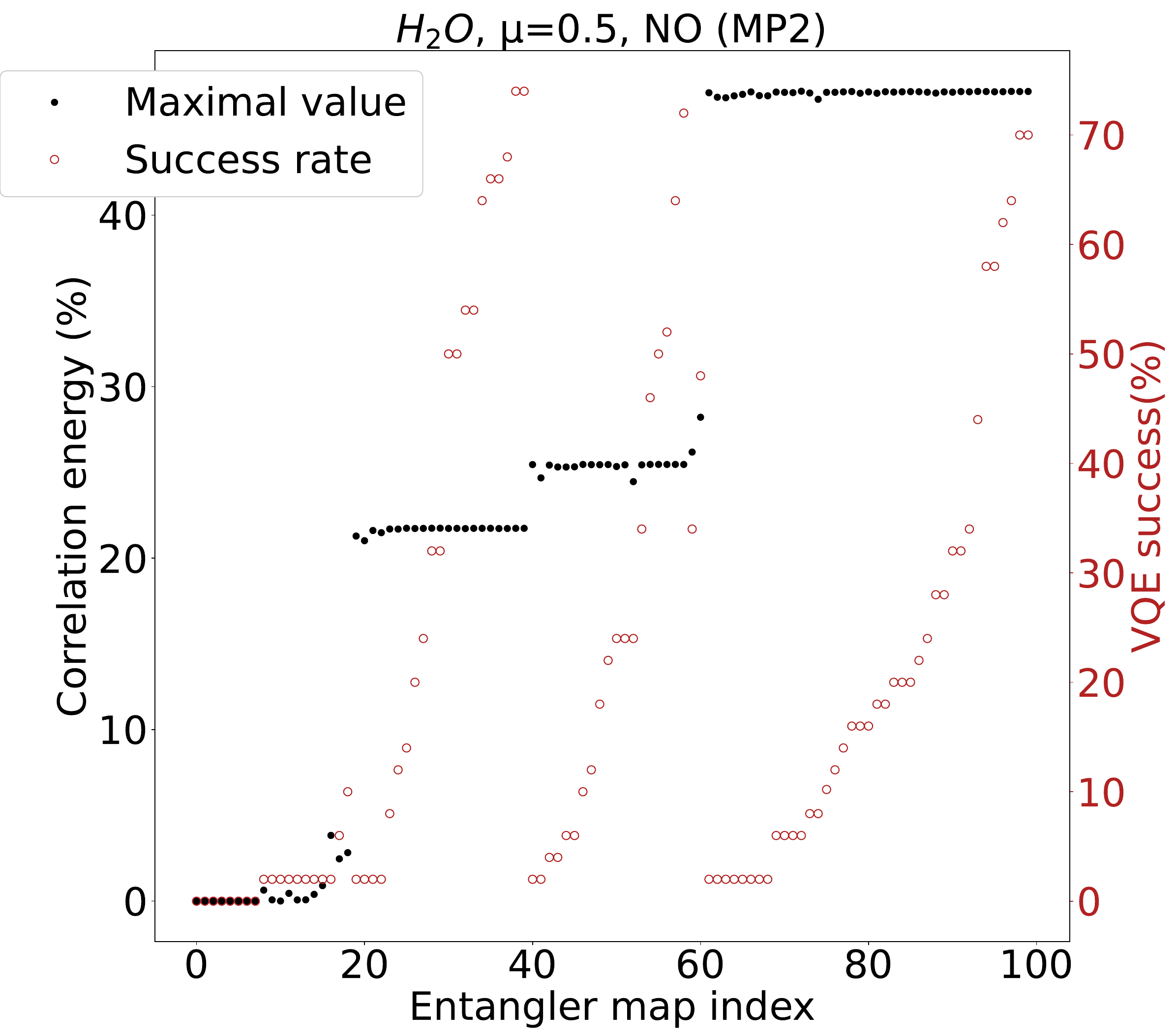}
    \caption{Multiple VQEs experiments for the $H_2O$ molecule with different ordering of the same parent sequence corresponding to $\mu=0.5$. Depth is equal to 2 for each ansatz generated by an entangler map. 50 VQEs runs are performed for each ansatz and the left vertical axis shows the maximum value of the percentage of the correlation energy while the right axes (red) represent the percentage of the 50 VQEs that are within 30\% of their respective maximum (black spot). We point out that the rising order in the representation has no objective meaning other than to facilitate the reader to identify the red-black dots pair.}
    \label{fig:singledepth}
\end{figure}

In this section, we will show the results obtained by applying the QIDA strategy described in the previous sections to the selected molecular systems of table \ref{tab: systems}. 
First of all, we notice that the use of the NO (see Sec. ~\ref{subsec:NO}) is crucial to increase the sparsity of the $I$ matrix, i.e. to reduce as much as possible the number of qubits pairs with significant values of the QMI.
This concept is illustrated in figure \ref{fig:maps}: for each molecule under investigation, we have two maps corresponding to the FCI state expressed in the HFCO and NO basis, respectively on the left and on the central columns. It is clear that the use of the NO is optimal to be used in combinantion with empirical ans\"atze since it increases
the sparsity of the QMI matrix, reducing the circuit depth, as already pointed out by our previous work \cite{Ratini2023}. By comparing the images in the central column with those in the right column, one notes the close similarity of the QMI matrices obtained from the FCI and the MP2 states expressed in NO (a similar behavior is also claimed for DMRG calculation \cite{Rissler2006}), giving consistency to the choice of MP2 state as for input orbitals and QIDA ans\"atze on the quantum circuit.
Regarding the impact of the different orderings of the parent sequences, we analyze their relative performance in figure \ref{fig:singledepth}, where we report the results for the $H_2O$ molecule.
By setting $\mu=0.5$, we generated 100 different entangler maps from the parent sequence. Each one of them is used as a different ansatz with depth 2, as reported on the horizontal axis of figure \ref{fig:singledepth}. For each ansatz we performed 50 independent VQEs run and reported, on the left vertical axis, the maximal value of the percentage of the correlation energy achieved. The right vertical axis gives the probability that the energy of one of the 50 VQEs lies within 30\% of the maximum, the value shown on the left vertical axis. Please note that the ordering on the basis of vertical left axes values is not meaningful of any intrinsic property, it was only imposed to facilitate the visual pairing between black and red spots.
The data clearly shows how, as expected, some entangling maps perform statistically better than others.

\begin{figure*}
    \centering \includegraphics[width=0.8\textwidth]{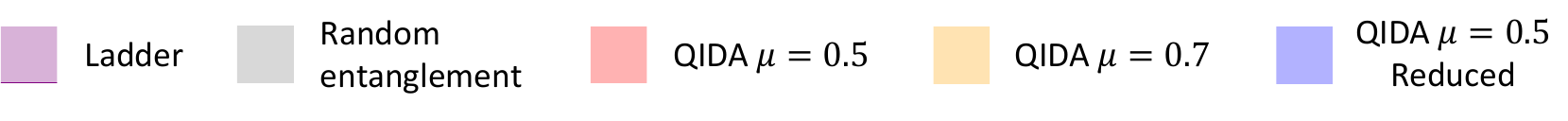}
    \\[\smallskipamount]
    \includegraphics[width=0.45\textwidth]{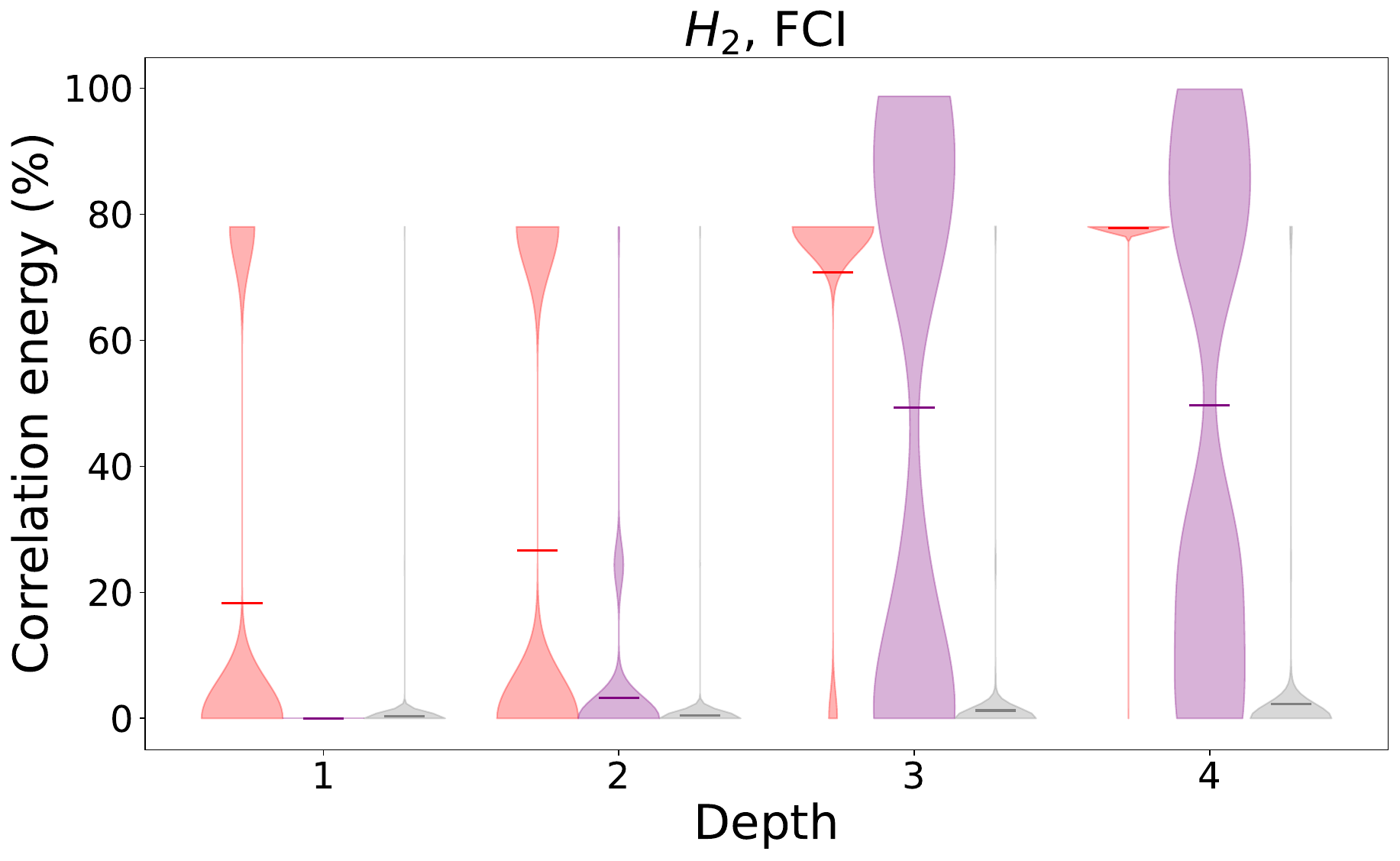}\hfill
    \includegraphics[width=0.45\textwidth]{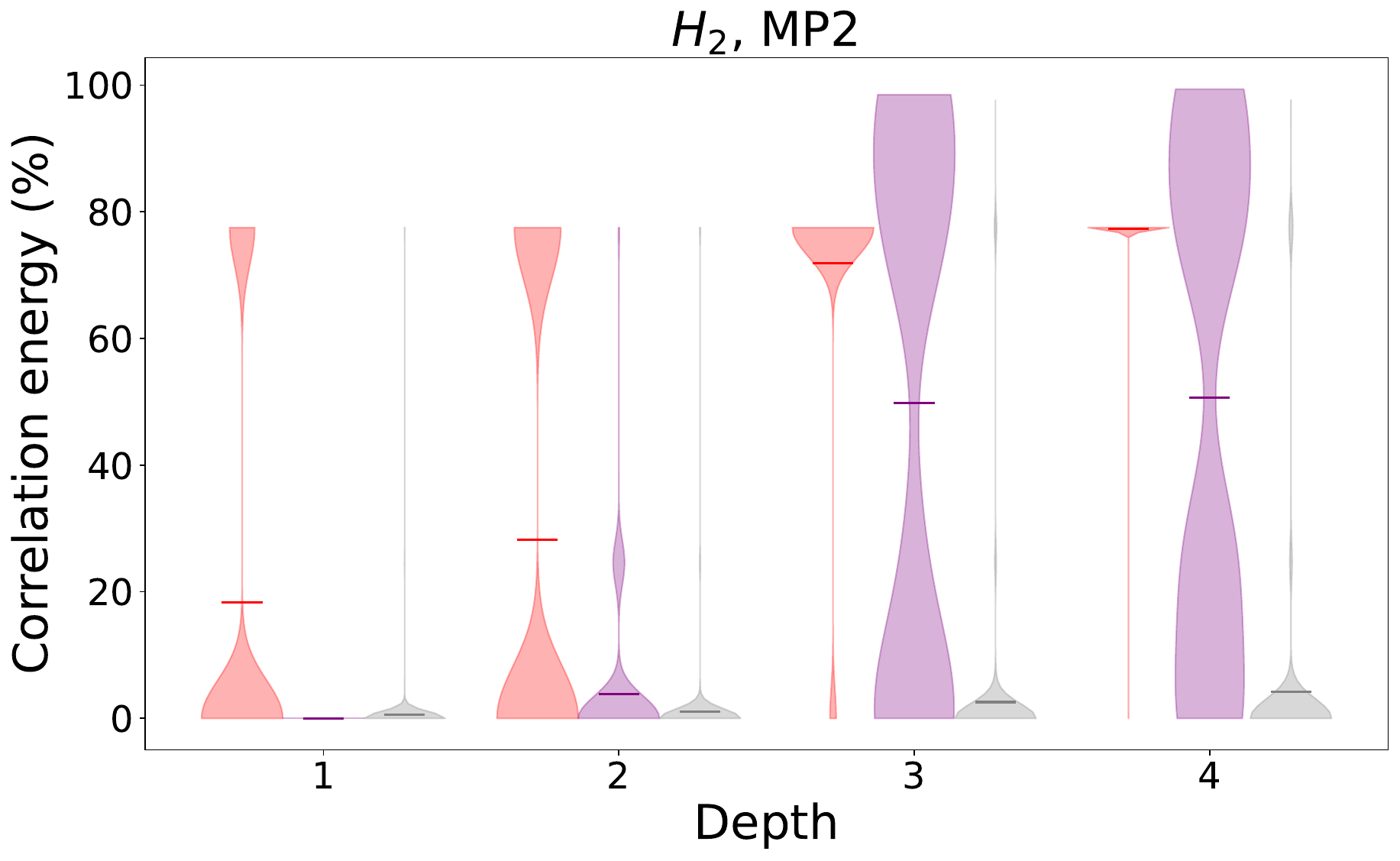}
    \\[\smallskipamount]
    \includegraphics[width=0.45\textwidth]{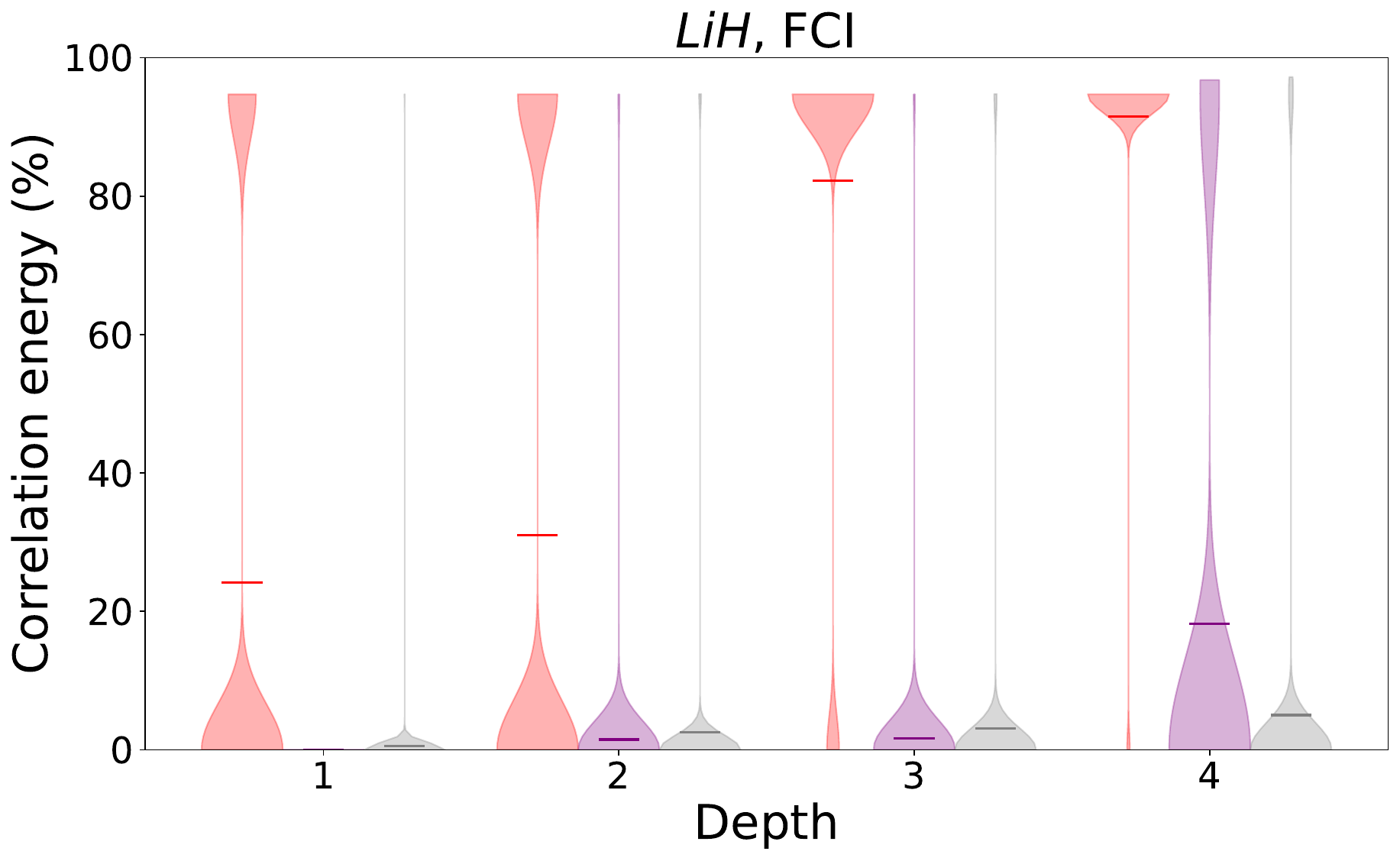}\hfill
    \includegraphics[width=0.45\textwidth]{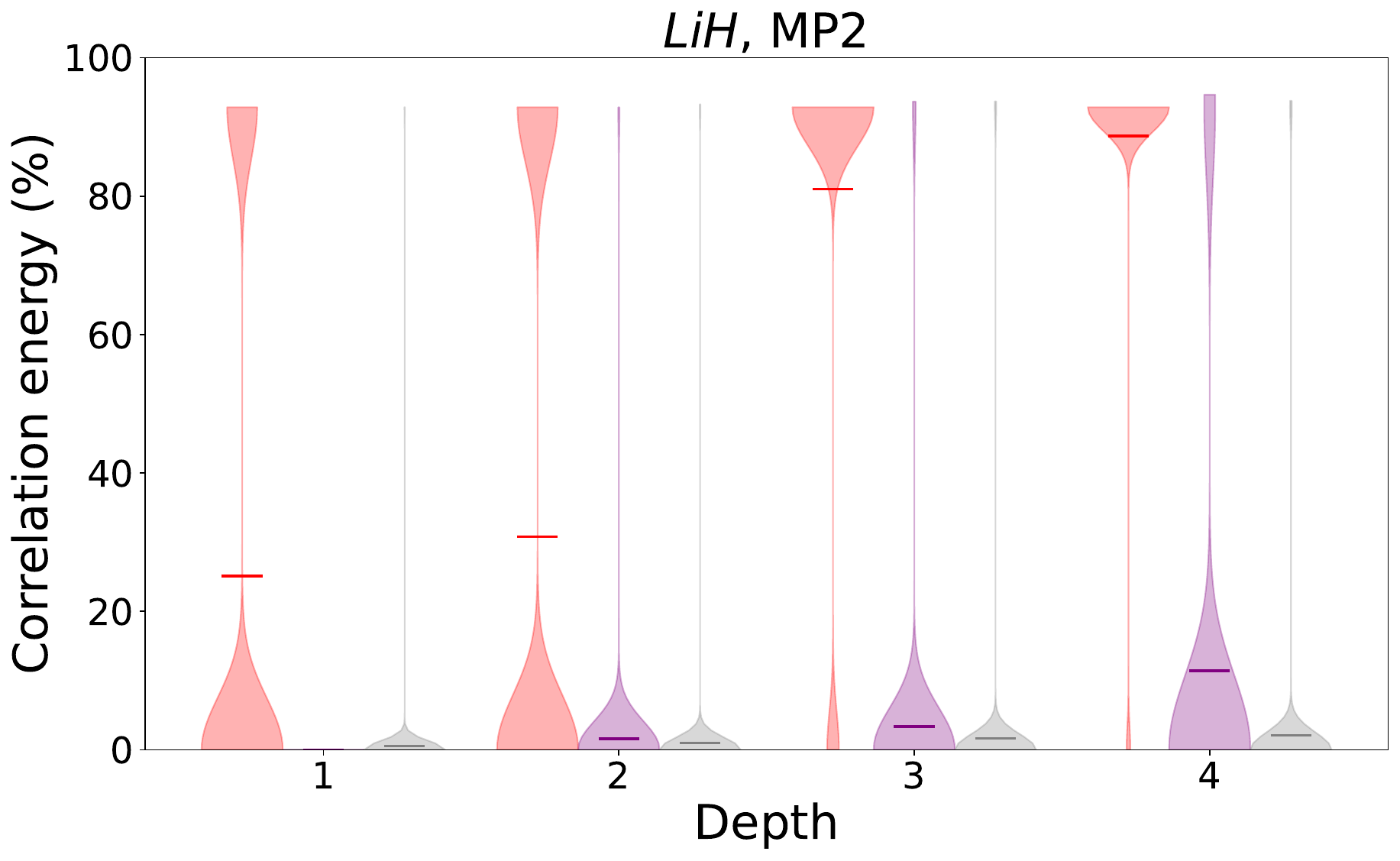}
    \\[\smallskipamount]
    
    \includegraphics[width=0.45\textwidth]{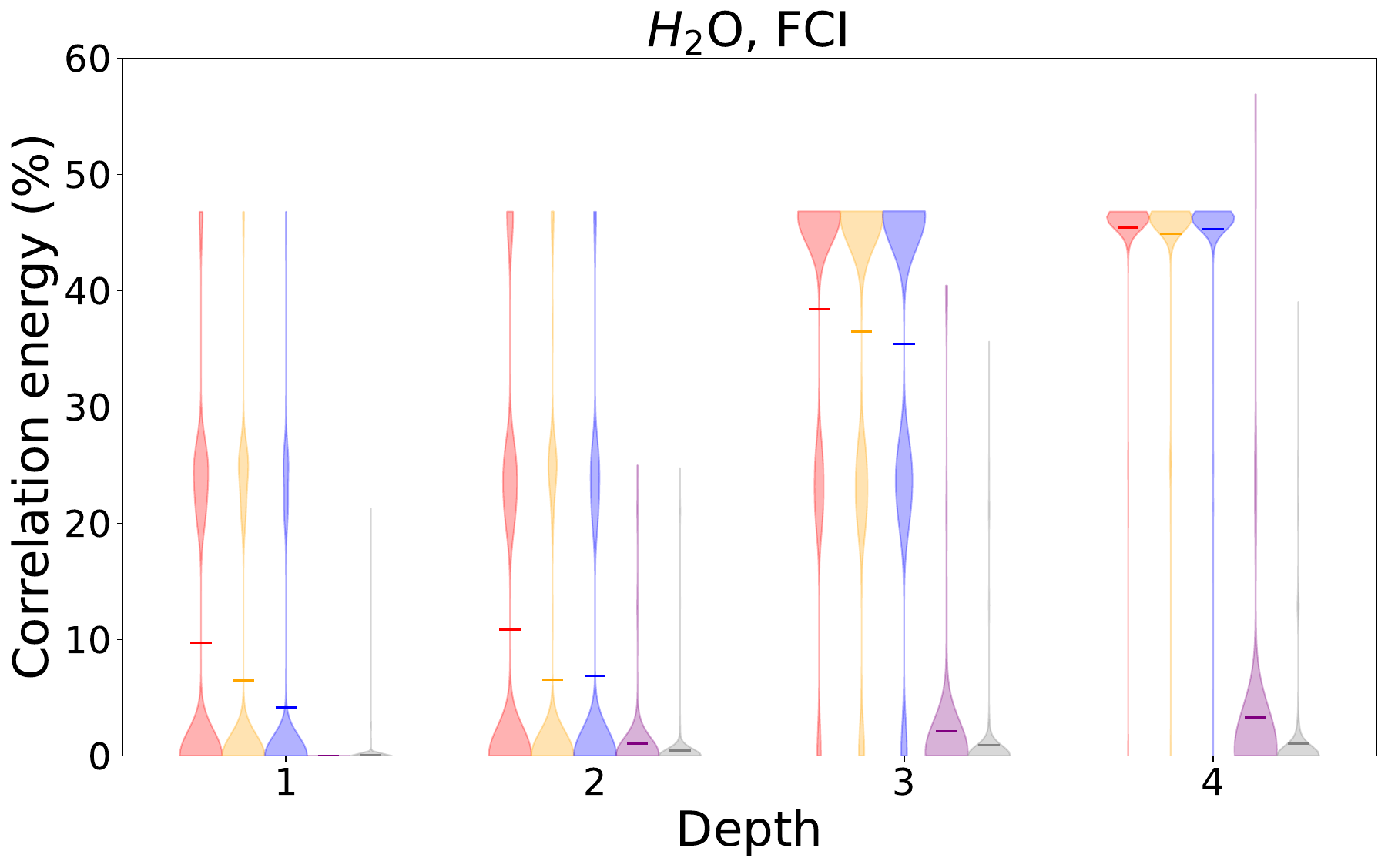}\hfill
    \includegraphics[width=0.45\textwidth]{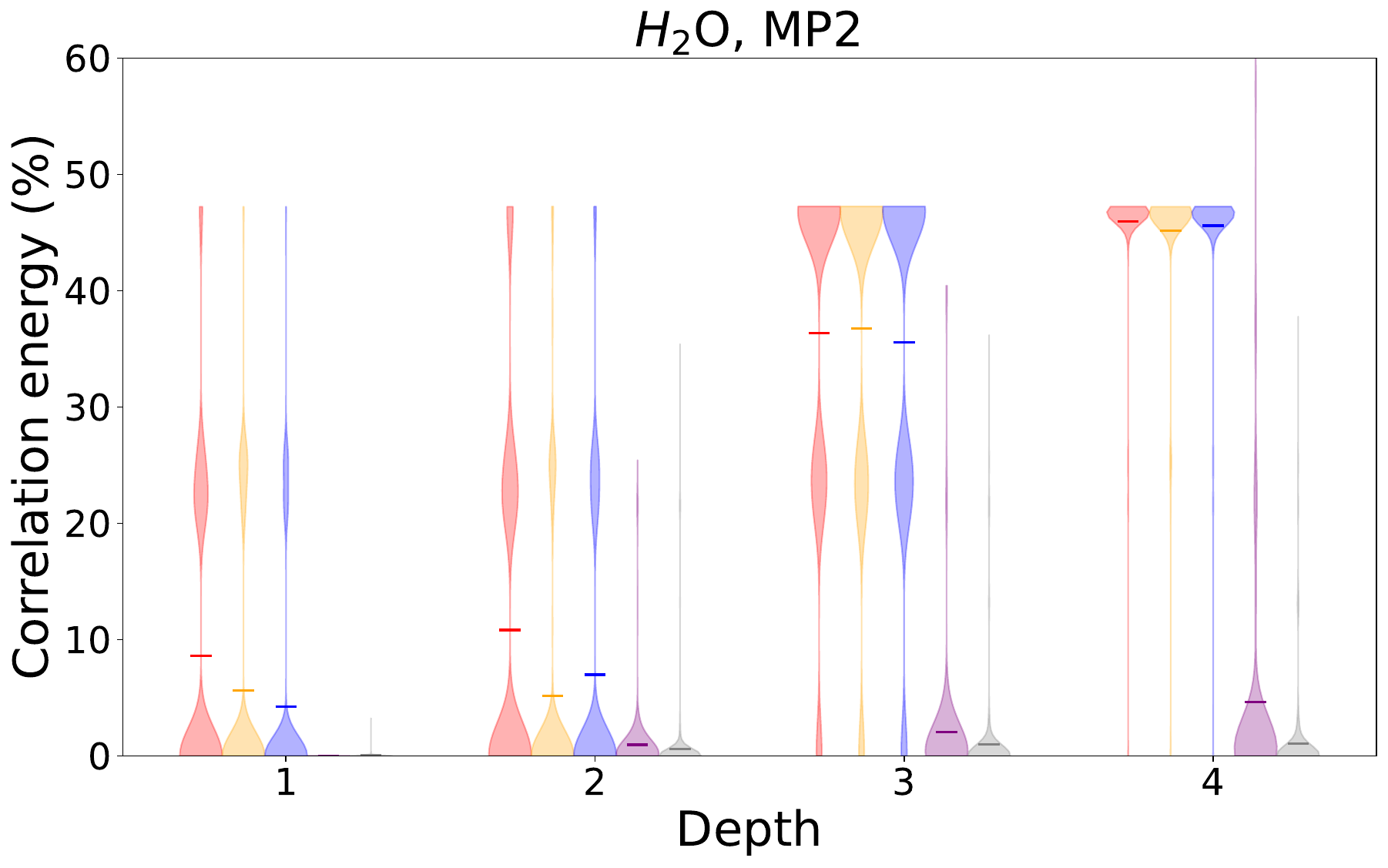}
    \caption{Average performance of our ans\"atze compared to the ladder-entangler and random entangler ones. Each violin column corresponding to a QIDA ansatz represents the total distribution of the energy results obtained by executing 50 VQEs runs for each considered block belonging to a parent sequence. For the ladder-entangler ans\"atze, the VQE algorithm has been repeated 300 times with the exception of the $H_2O$ system with depth 4 (repeated 2000 times). The violins are all normalized only between themselves to achieve a fixed maximal allowed width.}
    \label{fig:violins}
\end{figure*}

Figure \ref{fig:violins} is obtained by taking multiple statistical datasets such as the one in figure \ref{fig:singledepth} and transposing them over a single column, with this procedure being repeated for all the depths and ans\"atze shown. In each one of these plots, the $x$-axis  defines the depth under analysis, the molecule is pointed out in the title together with the classical method that gave the input $NO$ for the quantum circuit QMI map as in Figure \ref{fig:maps}, the kind of ans\"atze is then specified by their respective color.

In this way, we effectively estimate the probability distribution ($y$-axis) of the results obtained by using the QIDA method. Figure \ref{fig:violins} gives the opportunity to statistically characterize the quality of the results obtained from different parent sequences. 
For $H_2, LiH, H_2O$ molecules, we considered one or more values for the $\mu$ parameter that gives rise to different parent sequences, as reported in table \ref{tab: systems}, which also illustrates if the reduction explained in subsection \ref{subsubsec:reduction} has been taken into account. For each parent sequence, we extract 100 random entangling block orderings, as described in the subsection \ref{subsec:Ansätze}. These blocks are repeated to define ans\"atze with depths ranging from 1 to 4 and the corresponding VQE results are reported by the violin plots referred to as QIDA in figure \ref{fig:violins}. In such plots the width of the violin is proportional to the occurrences of the energies in the VQE statistics. The different kinds of QIDA ans\"atze are compared with the violins obtained from the random ans\"atze (details in subsection \ref{subsec:Ansätze}) and the ladder-entangler one over depths ranging from 1 to 4. 
Analyzing the dependence of the maximum value of the percentage of correlation energy on the circuit depth, we see that there is a plateau. Furthermore, already from depth 1 the results spread within the whole interval of the reachable values, with an average value of the correlation energy greater than the one of the ladder-entangler ansatz at depth 4.

Furthermore, we note that the maximal value of the percentage of the correlation energy is obtained by the ladder-entangler ansatz for depths equal to or larger than 3, for the $H_2$ molecule, and 4, for the $LiH$ and $H_2O$ molecules.  By comparing the mean values and the dispersion of the results around it, we see that QIDA ans\"atze assure an increased probability to obtain better approximations of the FCI state.
The rationale behind these results, and thus the explanation for the success of our method, lies in the information encoded in the entangled map which is based on MP2 calculations.
As the ladder-entangler ansatz does not include information about the specific problem for which it is being used, nor is based on any chemically motivated background, the resulting ansatz for the molecular system under investigation, will not necessarily be aimed at finding the optimal set of determinants of the ground state. 
In the proposed approach here, on the contrary, the topology of correlation is suggested by the precomputed approximated ground state through the QMI. This information directs the optimization towards the space with the expected correlation between qubits, facilitating the optimization and increasing the percentages of success in the VQE runs. 

In this regard, for the $H_2O$ molecule, the only system for which we analyzed different values of $\mu$, we did not observe any evidences of advantage in decreasing the value of $\mu$ from 0.7 to 0.5, besides a moderate decrease in the expected correlation energy value. This behavior can be explained considering that passing from $\mu=0.7$ to $\mu=0.5$ only the three CNOTs between qubits pair 2-8, 2-10, 4-8 are added and, following what has been explained in \ref{subsubsec:reduction}, these CNOTs can be considered redundant by the cross-entanglement analysis of the $H_2O$ molecule (fig~\ref{fig:maps}f).

The presented results can be compared with other approaches used to build heuristic ans\"atze, for example the approach used by Tkachenko et al. \cite{Tkachenko2021}, where results of VQE are reported for both the \textit{$H_2$} and \textit{LiH} molecules. In the case of \textit{$H_2$}  our QIDA ansatz with depth 1 (MP2) (composed of only 6 CNOTs and 12 parametrized gates) is reaching the value of $\%E_c=77.5$ with respect to their 
8-depth circuit, reaching $\%E_c=61,4$ (composed of 28 CNOTs and 64 parametrized $Ry$ quantum gates). 
Better statistical properties for the convergence of our QIDA ansatz requires a depth of 3 (composed of 18 CNOTs and 20 parametrized gates), which is still arguably a preferable choice.
For the case  of $LiH$ molecule, we note that percentage of correlation energy reported by Tkachenko2021 et al. \cite{Tkachenko2021} using a circuit with depth 8 (composed of 36 CNOTs and 82 $Ry$) is comparable with our QIDA ansatz since they reach approximately $92\%$ and $94\%$ of the total $E_c$, respectively.
Nevertheless, in the case of QIDA ansatz, the corresponding depth lies between 1 and 3, depending on the statistical distribution desired for the results (as shown in figure \ref{fig:violins}), with a number of CNOTs ranging from 6 to 18 and a number of $Ry$ quantum gates ranging from 14 to 22. 
In both cases it is clear how our approach is able to provide a more effective balance between circuit depth and quality of the obtained results, even when we requires a better statistical distribution.

\begin{wrapfigure}{l}{0.24\textwidth}
\includegraphics[width=0.23\textwidth]{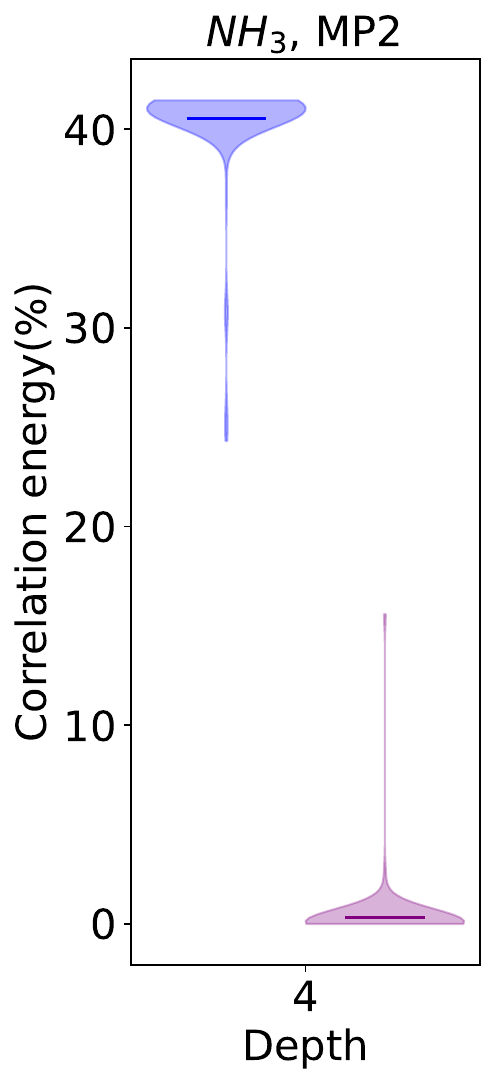}
 \caption{Statistical comparison of ladder-entangler ansatz (purple) with 100 repetitions and QIDA with 200 total VQEs run (blue).}
\label{fig:violinnh3}
\end{wrapfigure}

Every result shown up to now was simply to give a statistical analysis as extensive as possible to describe the effectiveness of the method described in this work. Having done that, we can now infer what behaviour to expect from other larger systems.\\
In figure \ref{fig:violinnh3} we reported the energy results obtained by applying the QIDA method (scheme in figure \ref{fig:scheme}) to the $NH_3$ molecule to compare the outcomes of the VQEs of our scheme with the ones obtained by the ladder-entangler map at depth 4. 
Thus, for this specific case, we tested our method by repeating 10 VQE runs for 18 different permutations of the parent sequence, while for the ladder entangling maps we performed 200 VQE runs.
Figure \ref{fig:violinnh3} shows an indisputable advantage for both the maximal value and statistical distribution of the converged VQEs energies. Furthermore, given the CNOTs connectivity of the ansatz, we achieve this result with a reduction in the number of CNOTs from 52 of the ladder-entangler ansatz to 36 of our method.
Finally, the trend shown by the molecules in figure \ref{fig:violins} suggests that the distribution of the reduced QIDA ansatz in figure \ref{fig:violinnh3} has already reached a plateau corresponding to the maximal value of the correlation energy percentage. Therefore, the ladder-entangler ansatz is still far from reliably finding even a moderate fraction of this quantity.

\begin{figure}[]
\includegraphics[width=0.45\textwidth]{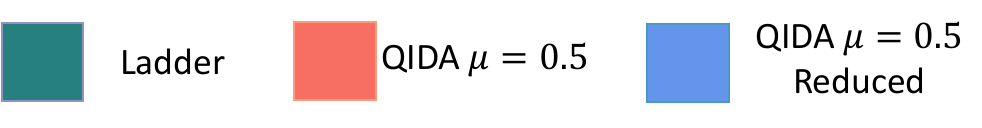}
\includegraphics[width=0.45\textwidth]{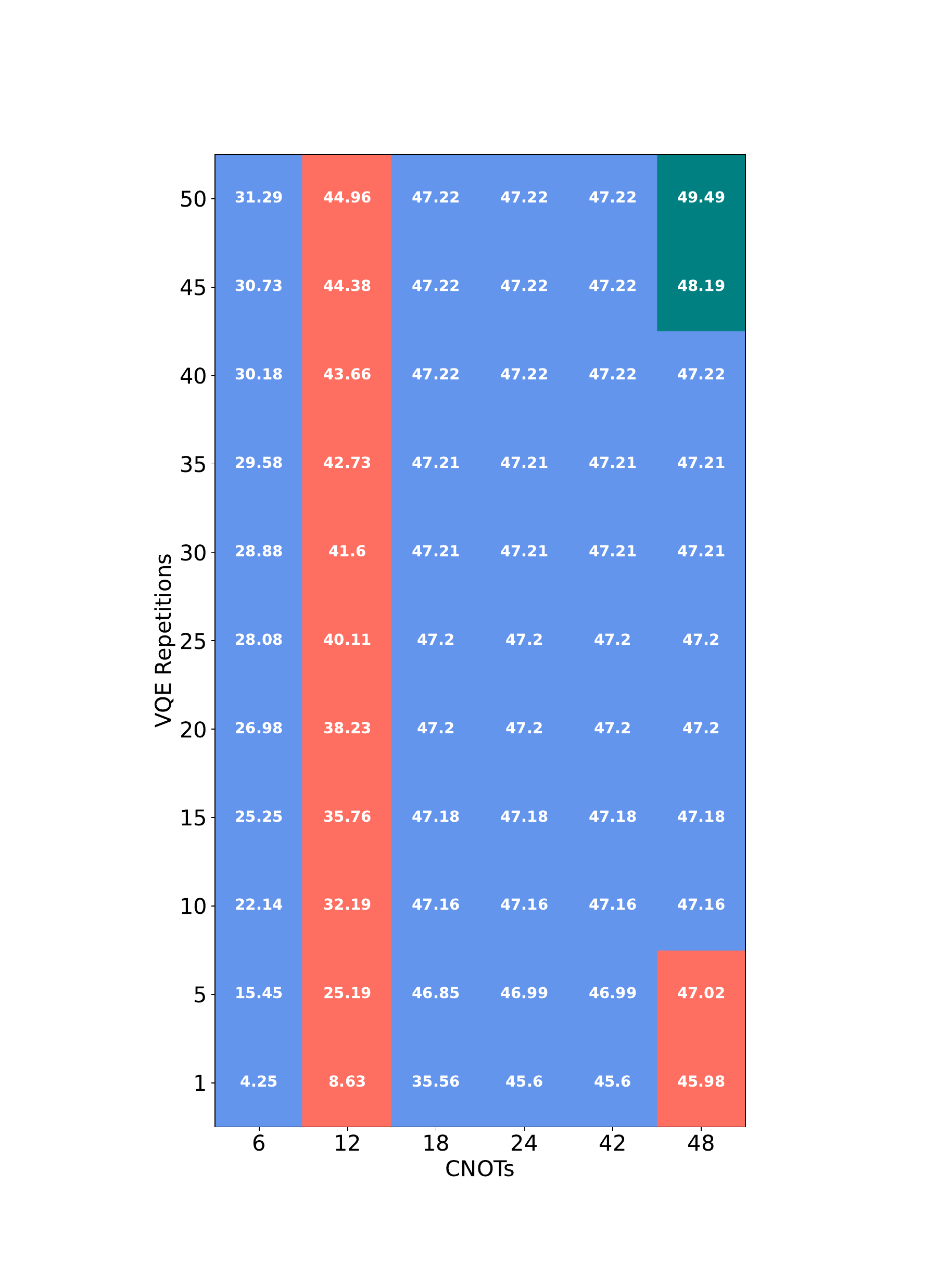}
\caption{Expected correlation energy as a function of resources: CNOTs(x axis) and VQE repetitions (y axis). The numerical values are the expectation values of the maximum of y VQE repetitions with an ansatz possessing no more than x CNOTs. The color of each spot is corresponding to the parent sequence that gives the highest results in terms of VQE correlation energy. The system under consideration is the $H_2O$ molecule.}
\label{fig:resources}
\end{figure} 
The previously discussed reduction of the number of needed CNOTs is also relevant for figure \ref{fig:violins}. For example, looking at figure \ref{fig:violins}[f], the reduced QIDA ansatz at depth 2 and the ladder-entangler ansatz at depth 4 have comparable statistical distributions of the energy results, with the QIDA ansanz having only 12 CNOTs versus 44 CNOTs of the ladder ansatz.
Moreover, these distributions have been obtained by repeating the VQE experiments many times. In terms of computational cost, the CNOTs and the number of VQE experiments can be seen as quantum resources, respectively x and y, and we want to investigate the relation between the energy results and these variables. Given a certain parent sequence $\sigma$, we introduce the quantity $\epsilon_{\sigma,\delta,i}$ that is the energy obtained at the $i-th$ VQE experiment by the ansatz composed of $\delta$ repetitions of $\sigma$, the sequence under investigation. Moreover, we introduce the function 
\begin{equation}
    f_{\sigma}(x)=max(\delta) \quad s.t. \quad CNOT(\sigma)*\delta\le x   
\end{equation}
which, fixed a sequence, finds the maximal number of repetitions that give rise to an ansatz such that the number of CNOTs is less or equal to $x$. Here, $CNOT(\sigma)$ gives the number of CNOTs corresponding to the sequence under consideration.
Finally, for each sequence $\sigma$, we define the function
\begin{equation}
   E_{\sigma}(x,y)=\braket{max_{k=1}^y max_{\delta=1}^{f_{\sigma}(x)}(\epsilon_{\sigma,\delta,k})}
   \label{eq:resources}
\end{equation}
that gives the expectation value of the maximal percentage of the correlation energy for each value of the computational resources $(x,y)$ introduced above.
As expected, for each pair $(x,y)$ the function in Eq.~\eqref{eq:resources} assumes different values by varying $\sigma$ and the maximal value is reported in figure \ref{fig:resources}. The associated spot has the colour of the corresponding $\sigma$, defined in figure \ref{fig:violins}.
The advantage of QIDA is remarkable in the overall results with respect to the ladder-entangler ans\"atz, particularly when the reduction is taken into account. When the number of CNOTs and VQE repetitions is increased, the ladder-entangler ansatz could give better maximal results, without yet increasing significantly the ones obtained with our method. 
This is not surprising because, as mentioned above, the former connects all the qubits and depends on more parameters, giving the possibility to catch part of the electronic correlations that the latter cannot reproduce even if the depth of the circuit is increased.
Overall, statevector simulations give the edge for the fully entangled ansatz only after many repetitions, however, if the noise of a real device was to be taken into account, we assert that the requirement of this ansatz for almost double the count of CNOTs with respect to the QIDA ansatz will likely degrade most of the advantage owned by it in the noiseless case.
Furthermore, these positive results encourage the idea of the QIDA ansatz as a modular starting circuit for state preparation that can be followed by a more complete fully entangled circuit that will be able to catch the missing correlations.

\section{\label{sec:conclusion}Conclusions}
In this paper, we have presented a novel approach to constructing hardware efficient ans\"atze for molecular systems. Our method leverages the quantum mutual information of the wavefunction of a well-scaling classical method, such as MP2, to generate ans\"atze for variational quantum algorithms that achieves correlation energies comparable or superior to currently used ladder-type heuristic ans\"atze, while requiring substantially fewer CNOTs.

To demonstrate the effectiveness of our approach, we studied various molecular systems using statevector simulations. 
For the $H_2O$ molecule, with just 24 CNOTs, our QIDA method achieved an expectation value of the correlation energy of one VQE of 45.56\%, significantly outperforming the ladder-entangler ans\"atz, requiring 44 CNOTs and yielding only 4.5\% of the correlation energy on average.
It is worth stressing that the results in figure \ref{fig:resources} have been obtained with statevector simulations. Then, for noisy simulations, we expect that QIDA ans\"atze will become even more efficient with respect to the ladder-entangler one because the number of two-qubits gates is significantly reduced.   
Additionally, we observed a growing performance disparity with the number of qubits (see fig.~\ref{fig:violinnh3}), with our QIDA method consistently demonstrating superior performances compared to the ladder-entangler ans\"atz.

Furthermore, we propose a modular approach to use our QIDA method in conjunction with other ans\"atze.
The idea is to first employ QIDA to identify the leading correlations between qubits in order to deliver optimized ans\"atze with essential albeit appropriate starting ingredients, excluding entanglement between loosely correlated qubits.
The state obtained running the VQE algorithm with such ansatz can be used as a starting point for subsequent application of entangling blocks that may refine the ansatz by catching the remaining correlation.
This approach has the potential to reduce the overall computational cost of quantum chemistry simulations while maintaining or improving the accuracy of the results.

In summary, our work has introduced a promising method for generating hardware-efficient ans\"atze for quantum chemistry inspired by quantum information. We have demonstrated the effectiveness of our QIDA method on a set of small molecules, increasing the percentage of the correlation energy values collected with fewer CNOTs compared to existing ansatz methods.

\subsection{Acknowledgments}
The authors acknowledge funding from the MoQS program, founded by the European Union’s Horizon 2020 research and innovation under the Marie Skłodowska-Curie grant agreement number 955479.
The authors also acknowledge funding from National Centre for HPC, Big Data and Quantum Computing - PNRR Project, funded by the European Union - Next Generation EU.\\
L.G. acknowledges funding from the Ministero dell'Università e della Ricerca (MUR) under the Project PRIN 2022 number 2022W9W423.

\newpage
\bibliographystyle{unsrt}
\bibliography{Cite}

\begin{thebibliography}{10}

\bibitem{Shor1997}
Peter~W Shor.
\newblock Polynomial-time algorithms for prime factorization and discrete
  logarithms on a quantum computer.
\newblock {\em SIAM Journal on Computing}, 26, 1997.

\bibitem{Egger2021}
Daniel~J. Egger, Claudio Gambella, Jakub Marecek, Scott McFaddin, Martin
  Mevissen, Rudy Raymond, Andrea Simonetto, Stefan Woerner, and Elena Yndurain.
\newblock Quantum computing for finance: State-of-the-art and future prospects.
\newblock {\em IEEE Transactions on Quantum Engineering}, 1:1--24, 1 2021.

\bibitem{Seskir2022}
Zeki~Can Seskir, Ramis Korkmaz, and Arsev~Umur Aydinoglu.
\newblock The landscape of the quantum start-up ecosystem.
\newblock {\em EPJ Quantum Technology 2022 9:1}, 9:1--15, 10 2022.

\bibitem{Troyer}
Vera von Burg, Guang~Hao Low, Thomas Häner, Damian~S Steiger, Markus Reiher,
  Martin Roetteler, and Matthias Troyer.
\newblock Quantum computing enhanced computational catalysis.
\newblock {\em Physical Review Research}, 3, 2021.

\bibitem{Robert2021}
Anton Robert, Panagiotis~Kl Barkoutsos, Stefan Woerner, and Ivano Tavernelli.
\newblock Resource-efficient quantum algorithm for protein folding.
\newblock {\em npj Quantum Information}, 7, 2021.

\bibitem{Feynman2018}
Richard~P Feynman.
\newblock Simulating physics with computers.
\newblock {\em Feynman and Computation}, 2018.

\bibitem{Peruzzo2014}
Alberto Peruzzo, Jarrod McClean, Peter Shadbolt, Man~Hong Yung, Xiao~Qi Zhou,
  Peter~J Love, Alán Aspuru-Guzik, and Jeremy~L O'Brien.
\newblock A variational eigenvalue solver on a photonic quantum processor.
\newblock {\em Nature Communications}, 5, 2014.

\bibitem{Bharti2022}
Kishor Bharti, Alba Cervera-Lierta, Thi~Ha Kyaw, Tobias Haug, Sumner
  Alperin-Lea, Abhinav Anand, Matthias Degroote, Hermanni Heimonen, Jakob~S.
  Kottmann, Tim Menke, Wai~Keong Mok, Sukin Sim, Leong~Chuan Kwek, and Alán
  Aspuru-Guzik.
\newblock Noisy intermediate-scale quantum algorithms.
\newblock {\em Reviews of Modern Physics}, 94:015004, 3 2022.

\bibitem{Cerezo2021}
M.~Cerezo, Andrew Arrasmith, Ryan Babbush, Simon~C. Benjamin, Suguru Endo,
  Keisuke Fujii, Jarrod~R. McClean, Kosuke Mitarai, Xiao Yuan, Lukasz Cincio,
  and Patrick~J. Coles.
\newblock Variational quantum algorithms.
\newblock {\em Nature Reviews Physics 2021 3:9}, 3:625--644, 8 2021.

\bibitem{Hoffmann1998}
Mark~R. Hoffmann and Jack Simons.
\newblock A unitary multiconfigurational coupled‐cluster method: Theory and
  applications.
\newblock {\em The Journal of Chemical Physics}, 88:993, 8 1998.

\bibitem{Kutzelnigg1991}
Werner Kutzelnigg.
\newblock Error analysis and improvements of coupled-cluster theory.
\newblock {\em Theoretica Chimica Acta}, 80:349--386, 7 1991.

\bibitem{Cooper2010}
Bridgette Cooper and Peter~J. Knowles.
\newblock Benchmark studies of variational, unitary and extended coupled
  cluster methods.
\newblock {\em The Journal of Chemical Physics}, 133:234102, 12 2010.

\bibitem{Evangelista2011}
Francesco~A. Evangelista.
\newblock Alternative single-reference coupled cluster approaches for
  multireference problems: The simpler, the better.
\newblock {\em The Journal of Chemical Physics}, 134:224102, 6 2011.

\bibitem{Whitfield2011}
James~D. Whitfield, Jacob Biamonte, and Alan Aspuru-Guzik.
\newblock Simulation of electronic structure hamiltonians using quantum
  computers.
\newblock {\em http://dx.doi.org/10.1080/00268976.2011.552441}, 109:735--750, 3
  2011.

\bibitem{Barkoutsos2018}
Panagiotis~Kl Barkoutsos, Jerome~F. Gonthier, Igor Sokolov, Nikolaj Moll, Gian
  Salis, Andreas Fuhrer, Marc Ganzhorn, Daniel~J. Egger, Matthias Troyer,
  Antonio Mezzacapo, Stefan Filipp, and Ivano Tavernelli.
\newblock Quantum algorithms for electronic structure calculations:
  Particle-hole hamiltonian and optimized wave-function expansions.
\newblock {\em Physical Review A}, 98:022322, 8 2018.

\bibitem{Romero2018}
Jonathan Romero, Ryan Babbush, Jarrod~R. McClean, Cornelius Hempel, Peter~J.
  Love, and Alán Aspuru-Guzik.
\newblock Strategies for quantum computing molecular energies using the unitary
  coupled cluster ansatz.
\newblock {\em Quantum Science and Technology}, 4:014008, 10 2018.

\bibitem{Kandala2017}
Abhinav Kandala, Antonio Mezzacapo, Kristan Temme, Maika Takita, Markus Brink,
  Jerry~M Chow, and Jay~M Gambetta.
\newblock Hardware-efficient variational quantum eigensolver for small
  molecules and quantum magnets.
\newblock {\em Nature}, 549, 2017.

\bibitem{rattew2020}
Arthur~G Rattew, Shaohan Hu, Marco Pistoia, Richard Chen, and Steve Wood.
\newblock A domain-agnostic, noise-resistant, hardware-efficient evolutionary
  variational quantum eigensolver.
\newblock {\em Preprint}, 2020.

\bibitem{Tang2021}
Ho~Lun Tang, V.~O. Shkolnikov, George~S. Barron, Harper~R. Grimsley,
  Nicholas~J. Mayhall, Edwin Barnes, and Sophia~E. Economou.
\newblock Qubit-adapt-vqe: An adaptive algorithm for constructing
  hardware-efficient ansätze on a quantum processor.
\newblock {\em PRX Quantum}, 2:020310, 4 2021.

\bibitem{Ratini2022}
Leonardo Ratini, Chiara Capecci, Francesco Benfenati, and Leonardo Guidoni.
\newblock Wave function adapted hamiltonians for quantum computing.
\newblock {\em Journal of Chemical Theory and Computation}, 18:899--909, 2
  2022.

\bibitem{mp2}
Chr Møller and M~S Plesset.
\newblock Note on an approximation treatment for many-electron systems.
\newblock {\em Physical Review}, 46, 1934.

\bibitem{Ding2021}
Lexin Ding, Sam Mardazad, Sreetama Das, Szilárd Szalay, Ulrich Schollwöck,
  Zoltán Zimborás, and Christian Schilling.
\newblock Concept of orbital entanglement and correlation in quantum chemistry.
\newblock {\em Journal of Chemical Theory and Computation}, 17:79--95, 1 2021.

\bibitem{Legeza2003}
Legeza and J.~Sólyom.
\newblock Optimizing the density-matrix renormalization group method using
  quantum information entropy.
\newblock {\em Physical Review B}, 68:195116, 11 2003.

\bibitem{Rissler2006}
Jörg Rissler, Reinhard~M. Noack, and Steven~R. White.
\newblock Measuring orbital interaction using quantum information theory.
\newblock {\em Chemical Physics}, 323:519--531, 4 2006.

\bibitem{Stein2016}
Christopher~J. Stein, Vera von Burg, and Markus Reiher.
\newblock The delicate balance of static and dynamic electron correlation.
\newblock {\em Journal of Chemical Theory and Computation}, 12:3764--3773, 8
  2016.

\bibitem{Tkachenko2021}
Nikolay~V Tkachenko, James Sud, Yu~Zhang, Sergei Tretiak, Petr~M Anisimov,
  Andrew~T Arrasmith, Patrick~J Coles, Lukasz Cincio, and Pavel~A Dub.
\newblock Correlation-informed permutation of qubits for reducing ansatz depth
  in the variational quantum eigensolver.
\newblock {\em PRX QUANTUM}, 2:20337, 2021.

\bibitem{Zhang2020}
Zi-Jian Zhang, Thi~Ha Kyaw, Jakob~S. Kottmann, Matthias Degroote, and Alán
  Aspuru-Guzik.
\newblock Mutual information-assisted adaptive variational quantum eigensolver.
\newblock {\em Quantum Science and technology}, 2021.

\bibitem{Grimsley2019}
Harper~R. Grimsley, Sophia~E. Economou, Edwin Barnes, and Nicholas~J. Mayhall.
\newblock An adaptive variational algorithm for exact molecular simulations on
  a quantum computer.
\newblock {\em Nature Communications}, 10, 12 2019.

\bibitem{Amico2008}
Luigi Amico, Rosario Fazio, Andreas Osterloh, and Vlatko Vedral.
\newblock Entanglement in many-body systems.
\newblock {\em Reviews of Modern Physics}, 80:517--576, 5 2008.

\bibitem{Neumann}
John von Neumann.
\newblock Mathematische grundlagen der quantenmechanik.
\newblock {\em Mathematische Grundlagen der Quantenmechanik}, 1996.

\bibitem{lowdin}
Per~Olov Löwdin and Harrison Shull.
\newblock Natural orbitals in the quantum theory of two-electron systems.
\newblock {\em Physical Review}, 101:1730, 3 1956.

\bibitem{Hartree1928}
D.~R. Hartree.
\newblock The wave mechanics of an atom with a non-coulomb central field. part
  ii. some results and discussion.
\newblock {\em Mathematical Proceedings of the Cambridge Philosophical
  Society}, 24:111--132, 1928.

\bibitem{Ratini2023}
Leonardo Ratini, Chiara Capecci, and Leonardo Guidoni.
\newblock Natural orbitals and sparsity of quantum mutual information.
\newblock {\em Preprint}, 2023.

\bibitem{Knowles1984}
P.~J. Knowles and N.~C. Handy.
\newblock A new determinant-based full configuration interaction method.
\newblock {\em Chemical Physics Letters}, 111:315--321, 11 1984.

\bibitem{Szabo1989}
Attila Szabo and Neil~S. Ostlund.
\newblock Modern quantum chemistry: Introduction to advanced electronic
  structure theory. revised.
\newblock {\em McGraw-Hili, New York}, pages 141--147, 1989.

\bibitem{Jordan1928}
P~Jordan and E~Wigner.
\newblock Über das paulische Äquivalenzverbot.
\newblock {\em Zeitschrift für Physik}, 47, 1928.

\bibitem{Bravyi2002}
Sergey~B Bravyi and Alexei~Yu Kitaev.
\newblock Fermionic quantum computation.
\newblock {\em Annals of Physics}, 298, 2002.

\bibitem{Parity}
Sergey Bravyi, Jay~M Gambetta, Antonio Mezzacapo, and Kristan Temme.
\newblock Tapering off qubits to simulate fermionic hamiltonians.
\newblock {\em Preprint}, 2017.

\bibitem{Zeng2019}
Bei Zeng, Xie Chen, Duan-Lu Zhou, and Xiao-Gang Wen.
\newblock {\em Quantum Information Meets Quantum Matter}.
\newblock Springer New York, 2019.

\bibitem{pyscf}
Qiming~Sun et~al.
\newblock Recent developments in the pyscf program package.
\newblock {\em The Journal of Chemical Physics}, 153:024109, 7 2020.

\bibitem{molcas}
Francesco Aquilante, Jochen Autschbach, Alberto Baiardi, Stefano Battaglia,
  Veniamin~A. Borin, Liviu~F. Chibotaru, Irene Conti, Luca~De Vico, Mickaël
  Delcey, Ignacio~Fdez Galván, Nicolas Ferré, Leon Freitag, Marco Garavelli,
  Xuejun Gong, Stefan Knecht, Ernst~D. Larsson, Roland Lindh, Marcus Lundberg,
  Per Åke Malmqvist, Artur Nenov, Jesper Norell, Michael Odelius, Massimo
  Olivucci, Thomas~B. Pedersen, Laura Pedraza-González, Quan~M. Phung,
  Kristine Pierloot, Markus Reiher, Igor Schapiro, Javier Segarra-Martí,
  Francesco Segatta, Luis Seijo, Saumik Sen, Dumitru~Claudiu Sergentu,
  Christopher~J. Stein, Liviu Ungur, Morgane Vacher, Alessio Valentini, and
  Valera Veryazov.
\newblock Modern quantum chemistry with [open]molcas.
\newblock {\em The Journal of Chemical Physics}, 152:214117, 6 2020.

\bibitem{Qiskit}
A.~Mitchell et~al.
\newblock Qiskit: An open-source framework for quantum computing, 2021.

\end{thebibliography}

\end{document}